\documentclass[10pt]{elsarticle}
\usepackage{epsfig}

\include{graphics}
\begin{document}

\title{Euler - Heisenberg effective action and magnetoelectric
effect in multilayer graphene}

\author[RUN]{M.I.~Katsnelson}

\author[AU,ITP]{G.E.~Volovik}

\author[ITEP]{M.A.~Zubkov}

\address[RUN]{Radboud University Nijmegen, Institute for Molecules
and Materials, Heyndaalseweg 135, NL-6525AJ Nijmegen, The Netherlands }

\address[AU]{Low Temperature Laboratory, School of Science and
Technology, Aalto University,  P.O. Box 15100, FI-00076 AALTO, Finland}

\address[ITP]{L. D. Landau Institute for Theoretical Physics,
Kosygina 2, 119334 Moscow, Russia}

\address[ITEP]{ITEP, B.Cheremushkinskaya 25, Moscow, 117259, Russia
}

\begin{abstract}
The low energy effective field model for the multilayer graphene (at ABC
stacking) is considered. We calculate the effective action in the presence of
constant external magnetic field $B$ (normal to the graphene sheet). We also
calculate the first two corrections to this effective action caused by the
in-plane electric field $E$ at $E/B \ll 1$ and discuss the magnetoelectric
effect. In addition, we calculate the imaginary part of the effective action in
the presence of constant electric field $E$ and the lowest order correction to
it due to the magnetic field ($B/E \ll 1$).
\end{abstract}



\maketitle

\section{Introduction}

Physics of graphene demonstrates numerous deep relations with fundamental
physics such as relativistic quantum mechanics and quantum field theory \cite%
{Katsbook}. The charge carriers in single-layer graphene are
topologically protected gapless fermions, which in the vicinity of the nodes
automatically acquire the properties of massless Dirac
fermions with relativistic spectrum. This naturally induces Lorentz boosts, e.g., in the problem of
graphene in crossed electric and magnetic fields \cite{Baskaran,Shytov}.
Bilayer \cite{bi} and rhombohedral (ABC-stacking) trilayer \cite{tri} graphene demonstrate
more exotic non-Lorentz-invariant physics of the topologically protected massless chiral fermions with
quadractic and cubic dispersion laws, respectively. Such theories are
intensively studied now, in particular, in a context of Ho\v{r}ava quantum
gravity with anisotropic scaling \cite{Horava} (this analogy was recently discussed in Ref. \cite%
{volovik2012}).  Quantum electrodynamics of bilayer graphene, which experiences
the phenomenon of anisotropic scaling, has been considered in Ref.
\cite{volovik2012} based on a semiclassical approach. Here we derive the
effective action of electromagnetic field in bilayer and multilayer graphene
based on exact solution of Schr\"odinger equation. For the case of purely
electric field, this has been done recently in Ref. \cite{zubkov}. We will
generalize this consideration to the case of crossed fields. This allows us to
discuss the magnetoelectric effect in graphene, that is, a change of
magnetization due to electric field and change of dielectric polarization due
to magnetic field.

In Ho\v{r}ava-Lifshitz  quantum
gravity  \cite{Horava}, the vacuum state  is characterized by anisotropic scaling, i.e. the action for gravity  is invariant under
transformation  ${\bf r}\rightarrow b {\bf r}$, $t\rightarrow b^z t$, where integer $z\neq 1$ is the analog of the dynamical critical exponent in the theory of phase transitions.  The Ho\v{r}ava-Lifshitz gravity in 3+1 spacetime is  asymptotic safe, i.e.  does not suffer from the ultraviolet divergences, if  $z=3$. It is instructive to extend the investigation of the consequences of the anisotropic scaling  to quantum field theories with general $z$. It appears that such theories are relevant for the models of the multilayer graphene. These models have  nodes in the fermionic spectrum, which are protected by the integer momentum-space topological invariant
$N$ expressed in terms of the Hamiltonian, see e.g. \cite{HeikkilaVolovik2010} (or more generally in terms of the Green's function \cite{Volovik2011}). Close to such a node fermions behave as 2+1 massless Dirac particles with energy
spectrum ${\cal E}(p) =\pm v p^N$, which obeys  the anisotropic scaling with $z=N$. That is why one may expect that the effective action  for quantum field theories emerging in such systems
contains the terms obeying this scaling law, and we consider here such terms in the effective
Euler-Heisenberg action for the quantum electrodynamics of the multilayer graphene.

In the simple model with $J$ ABC-stacked graphene layers considered in Ref. \cite{HeikkilaVolovik2010}, the topological invariant coincides with the number of layers,  $N=J$. We
keep in mind such model, though our consideration can be easily extended to the more general cases with $N\neq J$. We considered the Euler-Heisenberg effective action $S_{eff}({\bf B},{\bf E})$ in two limiting cases,  electric field dominated $B << E$ and magnetic field dominated  $E << B$.
Our results for the effective electromagnetic action  with
$E << B$ in  the system with number of graphene layers in the range $1\le J \le 10$ are accumulated in
Table \ref{tableseff}.  This Table demonstrates some peculiar properties of the action. In particular, the $\sim E^2$ correction due to electric
field decreases quickly with the increase of the number of layers while the coefficient before the main  magnetic term increases. At some special value of $J$ (at $J=4$) the $\sim E^2$ correction to the effective action vanishes.  There are also some special values of $J$, at which the logarithmically divergent term  appears; these are  $J=2,6,10$.  This situation is very similar to what occurs in the relativistic theories, where the logarithmically divergent terms appear only for some special values of space dimension $D$ (see e.g. \cite{Visser}). This suggests specific regularities in the behavior of quantum field theories, if they are extended to general values of $D$ and $z$.

Table \ref{tableseff} demonstrates that the linear term in electric field arises in the action. It is related with the degeneracy of the first $J$ Landau levels at $E=0$ (they are all zero modes) \cite{Katsbook,KP2008}. It describes the linear Stark effect (similar to that in hydrogen atom, due to degeneracy of the states with different orbital quantum numbers), which reflects the spontaneous (broken symmetry) electric polarization emerging in the system. This term can be presented as a scalar product ${\bf E} \cdot {\bf P}$, where the vector ${\bf P}$ is directed along the spontaneous polarization, which in the presence of electric field
is oriented along the field. This term is responsible for the magnetoelectric effect in Eq. (\ref{MAGNETOELECTRIC}).

In the electric field dominated regime $B << E$, our main result is the expression for the imaginary part of the effective action and the corresponding Schwinger pair creation rate given by Eqs. (\ref{OmegaEB}) and (\ref{GammaEB}).

The paper is organized as follows. In Section 2 we consider the system in pure magnetic field. In Section 3 the system is considered in the presence of both electric and magnetic fields with $E << B$. In Section 4 the opposite case $B << E$ is considered. In Section 5 we end with the conclusions.

\section{The system in pure magnetic field}

The Euler-Heisenberg effective action $S_{eff}$ (and Lagrangian $L_{eff}$)  in the considered effective field theory of  multilayer graphene is given by the expression for the  partition function of the system in the presence of magnetic field $B$ and electric field $E$:
\begin{eqnarray}
Z(T) & = & \int d \bar{\psi} d \psi \, {\rm exp}\Bigl(i \int_0^T dt d^2 x
\bar{\psi} [i
\partial_t - {\cal H}[E,B,t] ] \psi  \Bigr) = e^{i T L_{eff}[E,B]}=  e^{i  S_{eff}[E,B]} \label{Z_10}\label{ftfrolich0}
\end{eqnarray}
Here $\psi$ is the fermion field (with hidden spin index $A = 1,2$ and the
index $a = 1,2$ that enumerates the number of $2$ - spinors), $\cal H$ is the one -
particle Hamiltonian in the presence of external electric and magnetic fields
$E$ and $B$,  $T \rightarrow \infty$ is time extent. The boundary conditions relate the values of fields at $t = 0$ and $t = T$. Usually, for the solution of stationary problems, the anti - periodic boundary conditions are implied for the fermion systems (see, for example, Refs.\cite{Montvay} or \cite{semiclass}). However, in the case of constant nonzero electric field the system is not stationary. We overcome this difficulty considering the path integral in moving reference frame, where electric field vanishes. In this reference frame the problem becomes stationary, and we apply the anti - periodic boundary conditions for the fermion fields (for the details see Appendix A).

\subsection{Schr\"odinger equation}

\begin{figure}[h]
\centering
\includegraphics[width=13cm]{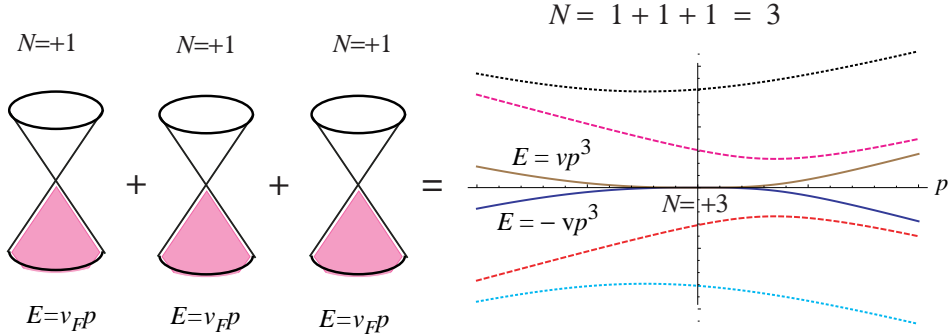}
\caption{
Illustration of branches of spectrum emerging in the system simulating the trilayer graphene with rhombohedral stacking
according to Ref. \cite{HeikkilaVolovik2010}. {\it left}: Noninteracting layers. Each layer has  its own
Dirac fermion with linear spectrum ${\cal E}(p) = \pm v_Fp$, (on this Figure $v_F$ is denoted by $c$, only one conical point at ${\bf p}=0$ is considered in each layer). The conical point is characterized by the momentum-space topological charge $N=1$, so that the total charge of the points at ${\bf p}=0$ is $N=3$ (for Dirac points in the other valley the charge is $N=-3$).
{\it right}: Interacting layers. The spectrum is shown as function of $p_x$ at $p_y=0$. Four branches of spectrum correspond to two massive Dirac fermions whose topological charge $N=0$, while
two branches with the lowest energy correspond to the massless Dirac particle with nonlinear
  dispersion  ${\cal E}(p) = \pm vp^3$, which is characterized by topological charge $N=3$.
  For the similar stacking of $J$ layers the branch of massless Dirac particles has
  dispersion  ${\cal E}(p) = \pm vp^J$, which corresponds to topological charge $N=J$.}
  \label{Cubic_spectrum}
\end{figure}

We start with the case of purely magnetic field. First, let us consider the
one-particle problem. Its solution is well known (see Chapter 2 in the book
\cite{Katsbook} and references therein) but it is convenient to repeat briefly
the main results for the case of $J$-layer graphene-like material with
arbitrary $J $. Here we consider an ideal system, in which the fermionic
spectrum is characterized by the nonlinear touching point ${\cal E}(p) = \pm
vp^J$ (see Fig. \ref{Cubic_spectrum} for $J=3$). It roughly corresponds to the
case of rhombohedral stacking of graphene layers in approximation, in which
only the largest in-plane and out-of-plane hopping parameters are taken into
account and the trigonal warping is neglected \cite{Katsbook}. The spectrum
with nonlinear touching points can be realized also in artificial materials as
it is done already for $J=1$ \cite{art1,art2}. It can also arise in
relativistic quantum field theories where the phenomenon of reentrant violation
of Lorentz symmetry takes place: the fundamental Lorentz violation at high
energies triggers  a reentrant violation of Lorentz invariance below some
low-energy scale leading to the non-linear touching point (see, e.g., Sec. 12.4
in the book~\cite{Volovik2003}). The Dirac vacuum of fermions, whose massless
branch of spectrum has non-linear touching points ${\cal E}(p) = \pm vp^J$,
induce the non-analytic terms in quantum electrodynamics, which obey the
anisotropic scaling ${\bf r} \rightarrow  a{\bf r}$, $t \rightarrow  a^J t$, i.e.
the magnetic and electric fields scale as $B \rightarrow Ba^{-2}$ and $E
\rightarrow Ea^{-(J+1)}$. Here we will be mainly interested in these
non-analytic terms, while the massive Dirac fields give rise to the
conventional analytic terms in the Euler-Heisenberg \cite{EH} action of the
type $\sim B^4/m^5$, where $m$ is the mass of Dirac particles on branches with
energy gap (see Eq. (46) of \cite{Visser}).
%

We deal with the two-component spinors describing electron states in graphene placed in the external magnetic field
$B$ directed along $z$-axis (normal to the graphene plane). We consider
the external vector potential in the form: $A_{y}=Bx$ (Landau gauge). The
one-particle Hamiltonian in a subsequent parametrization has the form \cite%
{falko,Multilayer1,Multilayer2}
\begin{equation}
H=v\left(
\begin{array}{cc}
0 & \Bigl(\hat{p}_{x}-i(\hat{p}_{y}+Bx)\Bigr)^{J} \\
\Bigl(\hat{p}_{x}+i(\hat{p}_{y}+Bx)\Bigr)^{J} & 0%
\end{array}%
\right)   \label{eq1}
\end{equation}%
Here $\hat{p}_{x}=-i\partial _{x}$, $v$ is a constant that is equal to Fermi
velocity $v_{F}$ for the case of monolayer, $1/2m$ ($m$ is the effective
mass) for the case of bilayer, etc.; we will use the units $\hbar =e=1$.
This Hamiltonian describes the lowest energy bands at $\left\vert \mathcal{E}%
\right\vert \ll \left\vert t_{\bot }\right\vert $, where $t_{\bot }$ is the
interlayer hopping parameter and the energy $\mathcal{E}$ is counted from
the band crossing point (neutrality point). Therefore $t_{\bot } \approx 0.4$ eV plays
the role of the ultraviolet (UV) cutoff energy (further we denote it by $\Lambda$)
for $J > 1$; for $J = 1$ the Dirac model is applicable till the energies of the order of
the in-plane nearest-neighbor hopping parameter $t \approx 3$ eV, which gives us the cutoff
energy for this case.
The values of $v$ and $\Lambda = t_{\bot}$ are related as $v \approx \Lambda^{1-J}v_F^J$.
From the opposite side of the small energies, the applicability
of the model (\ref{eq1}) is restricted by trigonal warping effects due to farther
interlayer hopping terms; the corresponding energy scale is about 1 to 10 meV \cite{Katsbook,falko,mayorov}.
For the case of bilayer, it was shown both experimentally \cite{mayorov} and theoretically
(see the recent paper \cite{honer} and references therein) that many-body effects can
result into a reconstruction of the ground state at these small energies. Since our model is,
anyway, inapplicable there we will assume an infrared cutoff (when necessary) of the order
of 1 to 10 meV. Many-body effects can also lead to a reconstruction of the ground state
for the case of rhombohedral (ABC) trilayer, see, e.g., Ref. \cite{HKV}. Detailed nature
of these instabilities is still unknown and is a subject of intensive debates. We will not take
into account the many-body effects in this paper restricting ourselves by the calculations
of the effective electromagnetic action in the approximation of noninteracting fermions only.

Stationary Schr\"odinger equation has the usual form
\begin{equation}
\mathcal{E}\Psi =H\Psi
\end{equation}%
Later on we imply periodic boundary conditions in space
coordinates. That is why $\Psi $ can be decomposed into the sum over the
quantized $y$ - momentum: $\Psi (t,x)=\sum\limits_{p_{y}}e^{ip_{y}y-i%
\mathcal{E}t}\psi _{p_{y},\mathcal{E}}(x)$. $\psi _{p_{y},\mathcal{E}}(x)$
is the eigenfunction of the Hamiltonian ($\mathcal{E}$ is the eigenvalue):
\begin{equation}
v\left(
\begin{array}{cc}
0 & \Bigl(\hat{p}_{x}-i({p}_{y}+Bx)\Bigr)^{J} \\
\Bigl(\hat{p}_{x}+i({p}_{y}+Bx)\Bigr)^{J} & 0%
\end{array}%
\right) \psi _{p_{y},\mathcal{E}}(x)=\mathcal{E}\psi _{p_{y},\mathcal{E}}(x)
\end{equation}

\subsection{Landau levels}
\label{SectLL}
Let us now introduce the notations:
\begin{equation}
\epsilon =\frac{\mathcal{E}}{vB^{J/2}},\quad u=\sqrt{B}(x+p_{y}/B)
\end{equation}%
Then
\begin{eqnarray}
(-i)^{-J}\epsilon {\psi }_{1} &=&[\partial _{u}+u]^{J}\psi _{2}  \nonumber \\
(-i)^{-J}\epsilon {\psi }_{2} &=&[\partial _{u}-u]^{J}\psi _{1}, \label{psi10}
\end{eqnarray}%
where $\psi _{p_{y},\mathcal{E}} = (\psi_1, \psi_2)^T $. For $\psi _{1,2}$ we have:
\begin{eqnarray}
&&\psi _{1}=\frac{1}{\epsilon }(-i)^{J}[\partial _{u}+u]^{J}{\psi }_{2}
\nonumber \\
&&\epsilon ^{2}{\psi }_{2}=[\partial _{u}-u]^{J}[u+\partial _{u}]^{J}\psi
_{2}  \label{psi1_}
\end{eqnarray}%
As for the oscillator we introduce the annihilation and creation operators $%
\hat{a}=\frac{1}{\sqrt{2}}[u+\partial _{u}],\hat{a}^{\dag }=\frac{1}{\sqrt{2}%
}[u-\partial _{u}]$ and $\hat{N}=\hat{a}^{\dag }\hat{a}$. Then

\begin{eqnarray}
&&\psi _{1}=\frac{2^{J/2}}{\epsilon }(-i)^{J}\hat{a}^{J}{\psi }_{2}
\nonumber \\
&&\epsilon ^{2}{\psi }_{2}=2^{J}[\hat{a}^{\dag }]^{J}\hat{a}^{J}\psi
_{2}=2^{J}\hat{N}(\hat{N}-1)...(\hat{N}-J+1)\psi _{2}  \label{psi1__}
\end{eqnarray}%
The normalized solutions are
\begin{eqnarray}
\psi _{1} &=&\pm \frac{(-i)^{J}}{\sqrt{2}}\frac{B^{1/4}}{2^{n-J}(n-J)!\sqrt{%
\pi }}e^{-u^{2}/2}H_{n-J}(u)  \nonumber \\
{\psi }_{2} &=&\frac{1}{\sqrt{2}}\frac{B^{1/4}}{2^{n}n!\sqrt{\pi }}%
e^{-u^{2}/2}H_{n}(u)  \label{psiH}
\end{eqnarray}%
with integer $n\geq J$ and the eigenvalues $\epsilon =\pm \sqrt{%
2^{J}n(n-1)...(n-J+1)}$. The solutions for $J>n\geq 0$ have $\psi _{1}=0$
and correspond to $\epsilon =0$:
\begin{eqnarray}
\psi _{1} &=&0  \nonumber \\
{\psi }_{2} &=&\frac{B^{1/4}}{2^{n}n!\sqrt{\pi }}e^{-u^{2}/2}H_{n}(u)
\label{psiH0}
\end{eqnarray}%
That is why at each value of $p_{y}$ we have Landau levels with $\mathcal{E}=0
$ (the degree of degeneracy is $J$ per flux quantum) and the nondegenerate
levels
\[
\mathcal{E}=\pm vB^{J/2}\sqrt{2^{J}n(n-1)...(n-J+1)}\label{LandauLevels}
\]%
As usual, we need $-L_{x}/2<p_{y}/B<L_{x}/2,p_{y}=\frac{2\pi }{L_{y}}%
K_{y},K_{y}\in Z$, where $L_{x,y}$ is the linear size of the graphene sheet
in $x,y$ direction.

\subsection{Naive expression for the effective action}

For pure magnetic field at zero temperature the effective action can be written
as follows (for the definition of the effective action $S_{eff}$ and effective
Lagrangian $L_{eff}$ and their relation to the spectrum of one - particle
excitations see Appendix A)
\[
-S_{eff}=-L_{eff}T = -T\sum \,|\mathcal{E}|/2=-g_{s}g_{v}\frac{TL_xL_yB}{2\pi }%
v(2B)^{J/2}\sum_{n=J}^{\infty }\sqrt{n(n-1)...(n-J+1)}
\]%
Here $\sum \,|\mathcal{E}|$ is the sum of the energies over all energy levels;  $g_{s}=2$ and $g_{v}=2$ are spin and valley degeneracies, $T$ is the period of the motion. This sum is
formally divergent. Using the Hawking trick of the zeta-function regularisation \cite{Hawking1977}
we represent it formally for $J=1$ through the $\zeta$-function:
\begin{equation}
-S_{eff}=-g_{s}g_{v}\frac{TL_xL_yB}{2\pi }v_{F}(2B)^{1/2}\zeta (-1/2)=g_{s}g_{v}%
\frac{TL^{2}B}{8\pi ^{2}}v_{F}(2B)^{1/2}\zeta (3/2)
\end{equation}
where we use the Riemann identity \cite{WW,Bateman}%
\begin{equation}
2^{1-s}\Gamma \left( s\right) \zeta \left( s\right) \cos \left( \pi
s/2\right) =\pi ^{s}\zeta \left( 1-s\right)
\end{equation}

For $J=1$ the effective action has to be Lorentz invariant. This means that
constant magnetic and electric ($E$) fields must enter via the combination $%
B^{2}-E^{2}/v_{F}^{2}$. Thus, we arrive at the effective action in the
presence of both fields:
\begin{equation}
-S_{eff}=\frac{1}{\sqrt{2}}g_{s}g_{v}\frac{T L_x L_y}{4\pi ^{2}}
v_{F}[B^{2}-E^{2}/v_{F}^{2}]^{3/4}\zeta (3/2)
\end{equation}
This gives us a contribution of vacuum fluctuations (Euler-Heisenberg action
\cite{EH}) for the single-layer graphene.

From here we get at $B=0$ the imaginary effective action which describes a
spontaneous creation of electron-hole pairs from vacuum (Schwinger effect
\cite{Schwinger1951,SV1992,gavrilov96,ACM,cohen2008,Vildanov2009,Fialkovsky:2011wh,katsnelson2011}):
\begin{equation}
-2\mathrm{Im}S_{eff}=g_{s}g_{v}\frac{TL_xL_y}{4v_{F}^{1/2}\pi ^{2}}E^{3/2}\zeta
(3/2)
\end{equation}
This result coincides with that obtained for the vacuum persistence
probability using the other methods (see Refs.\cite{ACM,volovik2012} and
references therein).

\subsection{Zeta function regularization of the partition function}

Let us evaluate the Euler-Heisenberg effective action using the zeta-function
regularization \cite{Elizalde}. As it is shown in Appendix D this
regularization gives the correct values of the Euler-Heisenberg effective
action if the divergences are at most logarithmic. In the presence of the
divergences that are stronger than logarithmic the zeta - regularized
expression for the effective action gives subdominant terms.

To use this regularization we need to calculate the determinant of a positive
defined operator. Let us use the valley degeneracy for this purpose. We have
\begin{equation}
\Bigl(\mathrm{Det}\,[i\partial _{t}-\mathcal{H}]\Bigr)^{2}=\mathrm{Det}\,%
\Bigl(i\partial _{t}-\mathcal{H}\Bigr)\Bigl(\sigma _{3}[i\partial _{t}-%
\mathcal{H}]\sigma _{3}\Bigr)=\mathrm{Det}\,[i\partial _{t}-\mathcal{H}%
][i\partial _{t}+\mathcal{H}]=\mathrm{Det}\,[(i\partial _{t})^{2}-\mathcal{H}%
^{2}]
\end{equation}%
After the Wick rotation we get
\begin{equation}
\Bigl(\mathrm{Det}\,[\partial _{\tau }-\mathcal{H}]\Bigr)^{2}=\mathrm{Det}\,%
\Bigl(-[(i\partial _{\tau })^{2}+\mathcal{H}^{2}]\Bigr)
\end{equation}%
Zeta-function regularization \cite{Elizalde} then reads:
\begin{equation}
-L_{eff}=-{\cal T} \,\mathrm{Tr}\,\mathrm{log}\,\Bigl(\lbrack \partial _{\tau }-\mathcal{%
H}]\Bigr)^{g_{s}g_{v}}=\frac{1}{2}g_{s}g_{v}\partial _{s}\Bigl(\frac{1}{%
\Gamma (s)}\,\int_{0}^{\infty }dtt^{s-1}\mathrm{Tr}e^{-[(i\partial _{\tau
})^{2}+\mathcal{H}^{2}]t}\Bigr)|_{s=0}\label{detzeta}
\end{equation}%
Here the system is considered in imaginary time $\tau \in [0, {\cal T}]$, and  ${\cal
T}=1/T \rightarrow 0$ is the inverse temperature. Below we omit mentioning the exact
meaning of $T$ (real or imaginary time). It will always be clear from the
context.

It is worth mentioning that the zero modes are to be omitted here. We have:
\begin{eqnarray}
-L_{eff} &=&\frac{L_{x}L_{y}}{4\pi T}g_{s}g_{v}B \, \partial _{s}\Bigl(\frac{1}{%
\Gamma (s)} \int_{0}^{\infty
}dtt^{s-1}\sum_{l,n}\,e^{-[(\frac{2\pi }{T}(l+1/2))^{2}+\mathcal{E}%
_{n}^{2}]t}\Bigr)|_{s=0}  \nonumber \\
&=&\frac{L_{x}L_{y}}{8\pi\sqrt{\pi }}g_{s}g_{v}B \, \partial_{s}\Bigl(\frac{1}{%
\Gamma (s)} \int_{0}^{\infty
}dtt^{s-3/2}\sum_{m,n}\,e^{-\frac{m^{2}T^{2}}{4t}-\mathcal{E}_{n}^{2}t}\Bigr)%
|_{s=0}
\end{eqnarray}%
At zero temperature in the sum over $m$ only the term with $m=0$ survives.
Zero modes of $\mathcal{H}$ are omitted and we get:
\[
-L_{eff}=\frac{L_{x}L_{y}}{4 \pi \sqrt{\pi }}g_{s}g_{v}B\,  \partial _{s}\Bigl(\frac{1}{%
\Gamma (s)} \int_{0}^{\infty
}dtt^{s-3/2}\sum_{n,\mathcal{E}_{n}>0}\,e^{-\mathcal{E}_{n}^{2}t}\Bigr)%
|_{s=0}
\]%
The last equation defines analytical function of $s$ at $s>s_{0}$ for some $%
s_{0}$. This function has to be continued to $s=0$.

Let us check ourselves considering the case $J=1$:
\begin{eqnarray}
-L_{eff} &=&\frac{L_{x}L_{y}}{4\pi T \sqrt{\pi }}g_{s}g_{v}B \,  \partial _{s}\Bigl(\frac{1}{%
\Gamma (s)} \int_{0}^{\infty
}dtt^{s-3/2}\sum_{n\geq 1}\,e^{-2v_{F}^{2}Bnt}\Bigr)|_{s=0}  \nonumber \\
&=&\frac{L_{x}L_{y}}{4\pi\sqrt{\pi }}g_{s}g_{v}B \, \partial _{s}\Bigl(\frac{1}{%
\Gamma (s)} \int_{0}^{\infty
}dtt^{s-3/2}\,\frac{1}{e^{2v_{F}^{2}Bt}-1}\Bigr)|_{s=0}  \nonumber \\
&=&\frac{L_{x}L_{y}}{4\pi\sqrt{\pi }}g_{s}g_{v}B\, \partial _{s}\Bigl(\frac{1}{%
\Gamma (s)}(2v_{F}^{2}B)^{1/2-s}%
\Gamma (s-1/2)\zeta (s-1/2)\Bigr)|_{s=0}  \nonumber \\
&=&\frac{L_{x}L_{y}}{4\sqrt{2}\pi ^{2}}g_{s}g_{v}v_{F}B^{3/2}\zeta (3/2)
\end{eqnarray}%
which coincides with the result obtained above.

For arbitrary $J$ we get:
\begin{eqnarray}
-L_{eff} &=&\frac{L_{x}L_{y}}{4\pi\sqrt{\pi }}g_{s}g_{v}B \, \partial _{s}\Bigl(\frac{1}{%
\Gamma (s)} \int_{0}^{\infty }dtt^{s-3/2}\sum_{n\geq
J}\,e^{-v^{2}(2B)^{J}\frac{n!}{(n-J)!}t}\Bigr)|_{s=0}
\nonumber \\
&=&\frac{L_{x}L_{y}}{4\pi\sqrt{\pi }}g_{s}g_{v}B \, \partial _{s}\Bigl(\frac{1}{%
\Gamma (s)}v^{1-2s}(2B)^{J/2-Js}%
\int_{0}^{\infty }duu^{s-3/2}\,\sum_{n\geq J}\,e^{-\frac{n!}{(n-J)!}u}\Bigr)%
|_{s=0}\label{LzetaE0}
\end{eqnarray}%
The resulting expression for the effective action is
\begin{eqnarray}
-L_{eff} &=&\frac{L_{x}L_{y}}{4\pi\sqrt{\pi }}g_{s}g_{v}2^{J/2}vB^{(J+2)/2} \, \partial _{s}\Bigl(\frac{1}{%
\Gamma (s)} v^{-2s} B^{-Js}%
\Gamma (s-1/2)f_{J}(s-1/2)\Bigr)|_{s=0},  \nonumber \\
&&f_{J}(s)=\frac{1}{\Gamma (s)}\int_{0}^{\infty }duu^{s-1}\,\sum_{n\geq
J}\,e^{-\frac{n!}{(n-J)!}u},\quad s>s_{0}\label{21}
\end{eqnarray}%
Here the function $f_{J}(s)$ is defined by the given integral for $%
s>s_{0}=1/J$ while its value at $s=-1/2$ enters the expression for the
effective action. The function $f_{J}(s)$ can be considered as the
generalization of Riemann zeta-function. For $s>1/J$ it is given by the series:
\begin{eqnarray}
f_{J}(s) &=&\sum_{n\geq J}\frac{1}{(n)_{-J}^{s}}  \nonumber \\
&&(n)_{-J}=n(n-1)...(n-J+1)\label{Riemann_f}
\end{eqnarray}

We have ($s\rightarrow 0$):
\begin{eqnarray}
f_{J}(s-1/2) &\approx & \frac{f_{J}^{(-1)}(-1/2)}{s} + f_J^{(0)}(-1/2)\label{fpole}
\end{eqnarray}

Procedure for the calculation of the values of $f_{J}^{(-1,0)}(- 1/2)$ is given
in Appendix C. It has been shown that $f_{J}^{(-1)}(- 1/2)$ may be nonzero for
$J = 2(2K+1), K\in Z$.

As a result for the values of $J\ne 2(2K+1)$   we obtain $L_{eff} = L_x L_y \,
l_{eff} $ with
\begin{eqnarray}
-l_{eff}&=& v B^{(J/2+1)}   \, \alpha_J,
\nonumber \\ \alpha_J &=&  \Bigl( \frac{1}{4\pi\sqrt{\pi }}g_{s}g_{v}2^{J/2}%
\Gamma (-1/2)f_{J}(-1/2)\Bigr),
 \quad
 J = 1,3,4,5,7,8,9,11,12,13,15,...\label{ds___odd_J0}
\end{eqnarray}%
Now using the data of Table \ref{tablefg} from Appendix C one can easily
calculate values of $\alpha_J$.

For $J = 2,6,10,14,...$ we have ($\frac{1}{\Gamma (s)}\approx s + \gamma s^2$):
\begin{eqnarray}
-l^{(0)}_{eff}&=&  v B^{(J/2+1)}   \, \alpha^{(-1)}_J\, {\rm log}\Bigl( \frac{2
v^2 B^J}
{\mu^2e^{r^{(0)}}}\Bigr),\nonumber  \\
&&r^{(0)} = \gamma+\psi(-1/2)+\frac{f^{(0)}_{J}(-1/2)}{f^{(-1)}_{J}(-1/2)} ,\nonumber\\
&&\alpha^{(-1)}_J = -\Bigl( \frac{1}{4\pi\sqrt{\pi }}g_{s}g_{v}2^{J/2}%
\Gamma (-1/2) f^{(-1)}_{J}(-1/2)\Bigr),\label{S26}
\end{eqnarray}
Here $\mu$ is a constant of the dimension of mass that is not fixed by the bare
theory. It is worth mentioning that $r^{(0)}$ can be absorbed by the
dimensional constant $\mu$ via its rescaling: $\tilde{\mu} = \mu
e^{r^{(0)}/2}$. Physically, the cutoff at $\mu$ originates from inapplicability of our model
at low energies due to trigonal warping and/or many-body effects as was discussed above.

\begin{table}
\begin{center}
\begin{small}
\begin{tabular}{|c|c|c|c|c|c|c|c|c|c|c|c|c|c|c|c|c|}
\hline
$J$ &{\bf  Effective lagrangian} $-l_{eff}/(g_sg_v)$   \\
\hline
$1$ & $0.04679084006 \times v_{F}[B^{2}-E^{2}/v_{F}^{2}]^{3/4}$ \\
\hline $2$ & $-   0.01989436789\,\times\,v B^{2} \, {\rm log}\Bigl( \frac{2 v^2
B^2}{\tilde{\mu}^2} \Bigr) -  0.1125395395 \times E B^{1/2} - 0.03490053651 \times \frac{E^{2}}{v B}$ \\
\hline $3$ & $ 0.2431821436 \times v_F^2 B^2 \Lambda^{-1} -0.07113483618\, \times v B^{5/2}- 0.1949242002\times E B^{1/2}- 0.02123563323\,\times \frac{E^{2}}{v B^{3/2}} $ \\
\hline $4$ & $0.2750863237 \times v_F^2 B^2 \Lambda^{-1}+ 0.1746037024\, \times v B^{3}-0.3462141635 \times E B^{1/2}$\\
\hline $5$ & $0.3484677731 \times v_F^2 B^2
\Lambda^{-1}+1.486414997 \times v B^{7/2} -0.4740836354\times E B^{1/2}  - 0.003908473257\,\times \frac{E^{2}}{v^{}B^{5/2}} $  \\
\hline $6$ & $ 0.4434404060 \times v_F^2 B^2 \Lambda^{-1} -
0.3133362943\,\times\, v B^{4} \,{\rm
log}\Bigl( \frac{2 v^2 B^4}{\tilde{\mu}^2}\Bigr)-0.6561212594\times E B^{1/2}  -0.001346430685\times \frac{E^{2}}{v B^{3}}  $  \\
\hline $7$ & $0.5560282734 \times v_F^2 B^2
\Lambda^{-1} + 7.837230303 \frac{B^4}{\Lambda^5}v_F^6 -3.222245513  v B^{9/2}-0.8183430244  E B^{1/2}  - 0.0004176771170 \frac{E^{2}}{v^{}B^{7/2}} $  \\
\hline $8$ & $  0.6849063582 \times v_F^2 B^2
\Lambda^{-1} +7.438809597\frac{B^4}{\Lambda^5}v_F^6  + 27.35355260 \times v B^{5}-1.026656793 \times E B^{1/2} - 0.0001187738012 \times \frac{E^{2}}{v^{}B^{4}} $  \\
\hline $9$ & $0.8295068879 \times v_F^2 B^2
\Lambda^{-1} + 8.754557174\frac{B^4}{\Lambda^5}v_F^6 +264.3063672  v B^{11/2} -1.217484080 E B^{1/2} - 0.00003135755874 \frac{E^{2}}{v^{}B^{9/2}} $ \\
\hline $10$ & $ 0.9895460087  v_F^2 B^2 \Lambda^{-1} +10.93814698
\frac{B^4}{\Lambda^5}v_F^6  -40.34826486     {\rm log}( \frac{2
v^2 B^{10}} {\tilde{\mu}^2 }) v B^{6}- 1.449119539\, E B^{1/2}   -0.775944555\,10^{-5}  \frac{E^{2}}{v B^{5}}   $ \\
\hline
\end{tabular}
\end{small}
\end{center}
\caption{Effective lagrangian $-L_{eff}/(L_x L_y  g_sg_v)$.  Here   $\Lambda$ is the ultraviolet cutoff ($v \approx \Lambda^{1-J}v_F^J$), while $\tilde{\mu}$ is the cutoff parameter of the regularization scheme. Both $\Lambda$ and $\tilde \mu$ depend on microscopic physics and should be considered as fitting parameters.
} \label{tableseff}
\end{table}

\subsection{Evaluation of the ultraviolet divergent terms}

Below we describe how the zeta - regularized effective action appears in a more
conventional expression for the fermionic determinant. Namely,
 we consider
\begin{eqnarray}
-L_{eff}&=&-\frac{g_{s}g_{v}}{2T}\mathrm{Tr}\,\mathrm{log}\,\Bigl(\frac{(i\partial _{\tau })^{2}+\mathcal{H}^{2}}{(i\partial _{\tau })^{2}+\mathcal{H}_0^{2}}\Bigr)^{} \nonumber\\
 &=& \frac{1}{2T}g_{s}g_{v} \int_{0}^{\infty }dtt^{s-1} \Bigl( \mathrm{Tr}e^{-[(i\partial
_{\tau })^{2}+\mathcal{H}^{2}]t}-\mathrm{Tr}e^{-[(i\partial _{\tau
})^{2}+\mathcal{H}_0^{2}]t}\Bigr)|_{s=0}\nonumber\\
&=&\frac{L_{x}L_{y}}{8\pi\sqrt{\pi }}g_{s}g_{v} \, \int_{0}^{\infty
}dtt^{s-3/2}\Bigl( 2B\, \sum_{n\geq
J}\,e^{-v^{2}(2B)^{J}\frac{n!}{(n-J)!}t}+JB-\int_0^{\infty} d \tau \,e^{-v^{2}
\tau^{J} t}\Bigr)|_{s=0}
\nonumber\\
&=&\frac{L_{x}L_{y}}{8\pi\sqrt{\pi }}g_{s}g_{v} \, v (2B)^{1+J/2}\,
\Gamma(s-1/2)\, F_J(s-1/2)|_{s=0}, \nonumber\\&& F_J(s) =
\frac{1}{\Gamma(s)}\int_{0}^{\infty }dtt^{s-1}\Bigl( \sum_{n\geq
J}\,e^{-\frac{n!}{(n-J)!}t}+J/2-\int_0^{\infty} d \tau \,e^{- \tau^{J}
t}\Bigr)\label{detusual}
\end{eqnarray}
Here ${\cal H}_0$ is the Hamiltonian of the system in the absence of external
magnetic field. It is worth mentioning that in this expression we do not omit
zero modes of the Hamiltonians $\cal H$ and ${\cal H}_0$ as these zero modes
are to cancel each other.

At $J=1$ Eq. (\ref{detusual}) is convergent, and we find numerically using
MAPLE package that $F(-1/2) = \zeta(-1/2)$. This means that for the case $J=1$
zeta - regularized effective action appears when the contribution of the free
fermions is subtracted. It is worth mentioning that in the resulting zeta -
regularized expression the zero modes of $\cal H$ are omitted while in Eq.
(\ref{detusual}) are not. This means that the zero mode contribution is
subtracted automatically when we subtract the contribution of free fermions.

For the cases $J\ge 2$ Eq. (\ref{detusual}) contains divergences:
\begin{eqnarray}
-L_{eff}&=&\frac{L_{x}L_{y}}{8\pi\sqrt{\pi }}g_{s}g_{v} \, v (2B)^{1+J/2}\,
\Gamma(s-1/2)\, F_J(s-1/2,\delta)|_{s=0}, \nonumber\\&& F_J(s,\delta) =
\frac{1}{\Gamma(s)}\int_{\delta}^{\infty }dtt^{s-1}\Bigl( \sum_{n\geq
J}\,e^{-\frac{n!}{(n-J)!}t}+J/2-\int_0^{\infty} d \tau \,e^{- \tau^{J} t}\Bigr)
= f_J(s) + F_J^{div}(s,\delta) \label{detusual_3}
\end{eqnarray}
Here $\delta = \frac{v^2(2B)^J}{\Lambda^2}$, where $\Lambda$ is the ultraviolet
cutoff.
 The divergent
terms $F_J^{div}(s,\delta)$ are calculated in Appendix D. It occurs that the
zeta regularized effective action at $J\ge 3$ appears as a subdominant term.
The values of the effective action for $J =1,...,10$ (together with the
corrections due to the small electric field) are presented in Table
\ref{tableseff}.  For $J=2$ the numerical factor in front of the term $\sim v
B^2\, {\rm log} \,\frac{\tilde{\mu}^2}{\Delta^2}$  given in Eq. (6), Eq. (8) of
\cite{volovik2012} is reproduced with the magnetic scale $\Delta = \sqrt{2} v
B$, see also Ref. \cite{Slizovskiy2012}. For $J\ge 3$ we keep only the dominant ultraviolet divergent terms
proportional to $B^2$ and the finite correction due to the electric field.

The constants in front of the ultraviolet divergent terms at $J\ge 3$ do not
have much sense because these terms appear to be of the same order as the terms
that come from the large values of energy ($>\Lambda$), where the energy
depends on the momentum (in the absence of external fields) as ${\cal E} \sim
|p|$. We do not evaluate the latter terms here.  According to
\cite{Katsbook,Multilayer1,Multilayer2} we have $v = \Lambda^{1-J} \, v_F^J$. Therefore,
the leading terms for $J \ge 3$ are $\sim v_F^2 B^2 \Lambda^{-1} $. The
complete effective action  with the subdominant non - analytical terms and the
divergent terms is given by  Eq. (27) with the values of $f_J(s)$ and
$F_J^{div}(s,\delta)$ is presented in Appendix C and Appendix D.

\section{Effective action in the presence of magnetic field and small
electric field}

\subsection{Schr\"odinger equation}

Let us now consider the case when weak in-plane electric field (along the $x$-axis) is added. We consider the external vector potential in
the form: $A_{x}=Et+By$. The one-particle Hamiltonian in a subsequent
parametrization has the form (cf. Eq.(\ref{eq1}))
\begin{equation}
H=v\left(
\begin{array}{cc}
0 & \Bigl((\hat{p}_{x}+Et+By)-i\hat{p}_{y}\Bigr)^{J} \\
\Bigl((\hat{p}_{x}+Et+By)+i\hat{p}_{y}\Bigr)^{J} & 0\label{H1}%
\end{array}%
\right)
\end{equation}%
Again, we try the wave function as $\Psi (t,x))=\sum_{p_{x}}e^{ip_{x}y}\psi
_{p}(t)$.

\subsection{Zero order approximation and Hall conductivity}

Similar to subsection \ref{SectLL} we change the variables:
\begin{equation}
\lambda =\frac{E}{\sqrt{2}vB^{(J+1)/2}},\quad u=\sqrt{B}(y+\frac{p_{x}+Et}{B}%
)
\end{equation}%
and denote
\begin{equation}
\hat{a}^{\dag }=\frac{1}{\sqrt{2}}[-\partial _{u}+u],\quad \hat{a}=\frac{1}{%
\sqrt{2}}[\partial _{u}+u]
\end{equation}%
Instead of Eq. (\ref{psi10}) the Schr\"{o}dinger equation reads
\begin{eqnarray}
i[\partial _{t}+vB^{J/2}\lambda (\hat{a}-\hat{a}^{+})]{\psi }_{1}
&=&v(2B)^{J/2}[\hat{a}^{+}]^{J}\psi _{2}  \nonumber \\
i[\partial _{t}+vB^{J/2}\lambda (\hat{a}-\hat{a}^{+})]{\psi }_{2}
&=&v(2B)^{J/2}[\hat{a}]^{J}\psi _{1}  \label{ut}
\end{eqnarray}%
When $E$ is small, in the zeroth order approximation we have the analogue of Eq. (\ref{psi1_}):
\begin{eqnarray}
i\partial _{t}{\psi }_{1} &=&v(2B)^{J/2}[\hat{a}^{\dag }]^{J}\psi _{2}
\nonumber \\
i\partial _{t}{\psi }_{2} &=&v(2B)^{J/2}[\hat{a}]^{J}\psi _{1}
\end{eqnarray}%
We have the same Landau levels as without electric field. However, the
centers of orbits $y_{c}=-[p_{x}+Et]/B$ move now slowly with the velocity $%
v_{c}=E/B$ along the $y$-axis. This means that the Hall effect takes place,
i.e. when electric field along $x$ axis is turned on, the current along $y$%
-axis appears if some of the energy states are occupied. The conductivity at
zero temperature can be calculated as

\[
\sigma =\frac{\bf J}{EL_{x}L_{y}}=g_{s}g_{v}\sum_{n}^{\infty }\frac{B}{2\pi E}%
\frac{E}{B}\theta (\mu -\mathcal{E}_{n})=\frac{g_{s}g_{v}}{2\pi }\Bigl(\frac{%
J}{2}+I(\mu )\Bigr)
\]%
Here $\mu >0$ is the chemical potential. It has been taken into account that
the zero energy levels are half-filled at $\mu =0$. The number of nonzero
Landau levels with $\mathcal{E}_{n}<\mu $ is denoted by $I(\mu )$. We come
to the well-known conclusion (see Ref.\cite{Katsbook}) that at $\mu =+0$
the Hall conductivity is equal to

\begin{eqnarray}
\sigma_{+0}&=& g_s g_v \frac{J}{4\pi}
\end{eqnarray}

\subsection{The first order correction}

In the next approximation we take into account the second term in the l.h.s. of
Eq. (\ref{ut}):
\begin{eqnarray}
&&[\epsilon +i\lambda (\hat{a}-\hat{a}^{\dag })]{\psi }_{1}=2^{J/2}[\hat{a}%
^{\dag }]^{J}\psi _{2}  \nonumber \\
&&[\epsilon +i\lambda (\hat{a}-\hat{a}^{\dag })]{\psi }_{2}=2^{J/2}[\hat{a}%
]^{J}\psi _{1}  \label{ute}
\end{eqnarray}%
or
\[
\epsilon \psi =\left(
\begin{array}{cc}
i\lambda (\hat{a}^{\dag }-\hat{a}) & 2^{J/2}[\hat{a}^{\dag }]^{J} \\
2^{J/2}[\hat{a}]^{J} & i\lambda (\hat{a}^{\dag }-\hat{a})%
\end{array}%
\right) \psi
\]%

Here ${\cal E} = v B^{J/2}\epsilon$ is the one - particle energy in the
reference frame moving with the velocity $E/B$. In zero order approximation its
value is given in subsection \ref{LandauLevels}.

One can easily see that the first-order term in expansion of $\epsilon $ in
powers of $\lambda $ vanishes for $n\ge J$. For $n\leq J-1$  there is the first
order correction to the energy for $J\ge 2$. This correction is given by the
eigenvalues $\chi^{(J)}_k \lambda$ of the $J\times J$ matrix $i \lambda \Omega$
with
\begin{equation}
\Omega = \left(\begin{array}{ccccc}0&1&0&...&...\\
-1 & 0 & \sqrt{2} &..&...\\
0&...&...& ...&...\\
0&...&-\sqrt{J-2}&0&\sqrt{J-1}\\
0&...&0&-\sqrt{J-1}&0
 \end{array}\right)
\end{equation}
In particular, we have:
\begin{eqnarray}
&&\chi^{(2)} = \pm 1;\nonumber\\
&&\chi^{(3)} = 0, \pm \sqrt{3}\nonumber\\
&&\chi^{(4)} = \pm 2.334414217, \pm 0.7419637843 \nonumber\\
&&\chi^{(5)} = 0, \pm 2.856970013, \pm 1.355626179\nonumber\\
&&\chi^{(6)} = \pm 3.324257431,\pm 0.6167065895, \pm 1.889175877\nonumber\\
&&\chi^{(7)} = 0, \pm 3.750439718, \pm 1.154405395, \pm 2.366759409 \nonumber\\
&&\chi^{(8)} = \pm 4.144547191,\pm 0.5390798102,\pm 1.636519041,\pm 2.802485863 \nonumber\\
&&\chi^{(9)} = 0, \pm 4.512745858,\pm 1.023255666,\pm 2.076847980,\pm 3.205429001 \nonumber\\
&&\chi^{(10)} = \pm 4.859462833, \pm 3.581823478, \pm 0.4849357082, \pm
1.465989092, \pm 2.484325840 \label{chi}
\end{eqnarray}

\subsection{The second order approximation}

One can easily see that the first-order term in expansion of $\epsilon $ in
powers of $\lambda $ vanishes for $n\ge J$ and the second-order term

\begin{equation}
\epsilon _{n}=\epsilon _{n}^{(0)}+\lambda ^{2}\sum_{k\neq n}\frac{1}{%
\epsilon _{n}^{(0)}-\epsilon _{k}^{(0)}}|\psi _{n}^{\dag }(\hat{a}^{\dag }-%
\hat{a})\psi _{k}|^{2}
\label{ute_2_1}
\end{equation}
should be considered. Here we use that at $|n|\geq J$
\begin{equation}
\psi _{n}=\frac{1}{\sqrt{2}}\left(
\begin{array}{c}
\frac{B^{1/4}}{2^{|n|}|n|!\sqrt{\pi }}%
e^{-u^{2}/2}H_{|n|}(u)  \\
\mathrm{sign}(n)\,\frac{B^{1/4}}{2^{|n|-J}(|n|-J)!\sqrt{%
\pi }}e^{-u^{2}/2}H_{|n|-J}(u)
\end{array}%
\right)
\end{equation}%
whereas at $|n|<J$
\begin{equation}
\psi _{n}=\left(
\begin{array}{c}
\frac{B^{1/4}}{2^{|n|}|n|!\sqrt{\pi }}e^{-u^{2}/2}H_{|n|}(u) \\
0%
\end{array}%
\right)
\end{equation}%

For $n\geq J$ Eq. (\ref{ute_2_1}) can be rewritten as
\begin{eqnarray}
\Delta \epsilon _{n} &=&\epsilon _{n}-\epsilon _{n}^{(0)}=\frac{\lambda ^{2}%
}{4}\Bigl(\frac{1}{\epsilon _{n}^{(0)}-\epsilon _{n+1}^{(0)}}[\sqrt{n+1}+%
\sqrt{n-J+1}]^{2}+\frac{1}{\epsilon _{n}^{(0)}-\epsilon _{n-1}^{(0)}}[\sqrt{n%
}+\sqrt{n-J}]^{2}  \nonumber \\
&&+\frac{1}{\epsilon _{n}^{(0)}+\epsilon _{(n+1)}^{(0)}}[\sqrt{n+1}-\sqrt{%
n-J+1}]^{2}+\frac{1}{\epsilon _{n}^{(0)}+\epsilon _{(n-1)}^{(0)}}[\sqrt{n}-%
\sqrt{n-J}]^{2}\Bigr)=\lambda ^{2}\eta _{n},\nonumber\\ && \eta_n =
\frac{(J-4)(2n-(J-1))}{2J (n)_{-J}^{1/2} 2^{J/2}} \label{ute_2_}
\end{eqnarray}%
In particular, in the case $J=1$ we have
\[
\epsilon _{n}-\epsilon _{n}^{(0)}=-\frac{3\lambda ^{2}}{\sqrt{2}}\sqrt{n}
\]
\qquad

For $n < J$ the second order correction vanishes.

In order to make the consideration of the one - particle spectrum complete, we
consider in Appendix B the problem in the gauge, where the electric field is
introduced via the scalar potential $A_0 = Ex$. In this gauge there are real
energy levels ${\cal E}_{n, p_y} = {\cal E}_n -\frac{E}{B} p_y$, where ${\cal
E}_n$ are calculated above while $p_y$ is the momentum along the $y$ axis. The
Galilean transformation of energy gives the values of the energy levels equal
to ${\cal E}_n$ in the reference frame moving along the axis orthogonal to $E$
with the velocity $E/B$.

\subsection{Effective lagrangian}

In principle, in the presence of external field $E$ the one - particle problem
is not stationary, and we do not have usual energy levels in the original
reference frame. However, we do have such levels in the frame moving with the
velocity $E/B$ along the axis orthogonal to the direction of electric field. In
this case the expression for the effective lagrangian  is derived in Appendix
A. It occurs that this expression coincides with the usual one $ \sum \, |{\cal
E}|/2 $, where the summation is over the energy levels of the system defined in
the moving reference frame. This allows us to calculate the effective
lagrangian in this case.

For $J=1$ the correction to the effective lagrangian reads:
\begin{equation}
-\Delta L_{eff}=\frac{BL_{x}L_{y}}{2{\pi }}g_{s}g_{v}v_{F}B^{1/2}%
\sum_{n=1}^{\infty }n^{1/2}\frac{3E^{2}}{2\sqrt{2}v_{F}^{2}B}=\frac{3}{4}%
\frac{E^{2}}{v_{F}^{2}B^{2}}L_{eff}^{(0)}
\end{equation}%
where we consider $\sum_{n=1}^{\infty }n^{1/2}$ as $\zeta (-1/2)$.

For $J\ge 2$ the dominant contribution is given by the term linear in $E$ due
to the splitting of the lowest Landau Level:

\begin{equation}
-\Delta L^{(1)}_{eff}=-\frac{BL_{x}L_{y}}{2{\pi }}g_{s}g_{v}v_{}B^{J/2}%
\sum_{n=0}^{J-1} (|\chi^{(J)}_n|/2) \frac{E}{\sqrt{2} v
B^{(J+1)/2}}=-\frac{L_{x}L_{y}T}{4 \sqrt{2} {\pi }}g_{s}g_{v} {E}{B^{1/2}}
\sum_{n=0}^{J-1} |\chi_n|
\end{equation}

For $J\ge 4$ the second order correction can be calculated without use of any
regularization:
\begin{eqnarray}
-\Delta L^{(2)}_{eff}&=&-\frac{BL_{x}L_{y}}{2{\pi }}g_{s}g_{v}v_{}B^{J/2}%
\sum_{n=J}^{\infty} \eta^{}_n \frac{E^2}{2 v^2 B^{J+1}}=-\frac{L_{x}L_{y}T}{4
{\pi }}g_{s}g_{v} \frac{E^2}{v B^{J/2}} \sum_{n=J}^{\infty}
\frac{(J-4)(2n-(J-1))}{2J (n)_{-J}^{1/2} 2^{J/2}}\nonumber\\
&=& -\frac{L_{x}L_{y}}{4 {\pi }}g_{s}g_{v} \frac{E^2}{v B^{J/2}}
\frac{(J-4)(2g_J(-1/2)-(J-1)f_J(1/2))}{2J  2^{J/2}}\label{Seff4}
\end{eqnarray}
Here in addition to the function $f_J$ given by Eq. (\ref{Riemann_f}) we use the function $g_J$. These functions are the generalizations of the Riemann zeta-function and are given by:
\begin{eqnarray}
f_{J}(s) &=&\sum_{n\geq J}\frac{1}{(n)_{-J}^{s}},\quad g_{J}(s)=\sum_{n\geq J}%
\frac{n}{(n)_{-J}^{s+1}},  \nonumber \\
&&(n)_{-J}=n(n-1)...(n-J+1)\label{fg}
\end{eqnarray}
For $J=4$ the correction in Eq. (\ref{Seff4}) vanishes. For $J>4$ expressions
of Eq. (\ref{fg}) are convergent.

In order to calculate the second order correction to the effective action at $J
= 1,2,3$ we need to apply a certain regularization. First, let us consider the
zeta - regularized expression. It  can be calculated as follows:
\begin{eqnarray}
 -\Delta L^{(2)}_{eff} &=&\frac{L_{x}L_{y}}{4\pi\sqrt{\pi }}g_{s}g_{v}B \, \partial _{s}\Bigl(\frac{1}{%
\Gamma (s)} \int_{0}^{\infty
}dtt^{s-3/2}\sum_{n\geq J}\,[e^{-[\Delta \mathcal{E}+v(2B)^{J/2}\sqrt{\frac{n!%
}{(n-J)!}}]^{2}t}-e^{-[v(2B)^{J/2}\sqrt{\frac{n!%
}{(n-J)!}}]^{2}t}]\Bigr)|_{s=0}   \nonumber\\
& =&-\frac{L_{x}L_{y}}{4\pi\sqrt{\pi }}g_{s}g_{v}B \, \partial _{s}\Bigl(\frac{1}{%
\Gamma (s)} v^{1-2s}(2B)^{J/2-Js}
\int_{0}^{\infty }duu^{s-3/2}\,\sum_{n\geq J}\,e^{-\frac{n!}{(n-J)!}u}u%
\frac{E^{2}}{v^{2}B^{J+1}}\eta _{n}\sqrt{\frac{(n)_{-J}}{2^{J}}}\Bigr)%
|_{s=0}  \nonumber \\
&=&- \frac{L_{x}L_{y}}{4\pi\sqrt{\pi }}g_{s}g_{v}B\, \partial _{s}\Bigl(\frac{1}{%
\Gamma (s)} v^{1-2s}(2B)^{J/2-Js}\nonumber\\&&
\int_{0}^{\infty }duu^{s-3/2}\,\sum_{n\geq J}\,e^{-\frac{n!}{(n-J)!}u}u%
\frac{E^{2}}{v^{2}B^{J+1}}\frac{2n(J-4)+(5J-J^{2}-4)}{2^{J+1}J}\Bigr)|_{s=0}
\end{eqnarray}%
This is the generalization of Eq. (\ref{LzetaE0}) to the case of nonzero electric field.
The sum over $n$ is convergent due to the exponential factor. Coefficients $%
\eta _{n}$ are defined in Eq. (\ref{ute_2_}). We obtain ($L_{eff} = L_x L_y  \,
l_{eff}$):

\begin{eqnarray}
-l^{(0)}_{eff}&=& v B^{(J/2+1)}  \, \partial _{s}  \Bigl(  \frac{1}{%
\Gamma (s)} v^{-2s}(2B)^{-Js} \frac{1}{4\pi\sqrt{\pi }}g_{s}g_{v}2^{J/2}%
\Gamma (s-1/2)f_{J}(s-1/2)\Bigr),  \\
-\Delta l^{(2)}_{eff} &=& v^{-1} \frac{E^{2}}{B^{J/2}} \partial _{s}  \Bigl( \frac{1}{%
\Gamma (s)} v^{-2s}(2B)^{-Js} \frac{1}{4\pi\sqrt{\pi }}g_{s}g_{v}2^{J/2}%
\Gamma (s+1/2)\nonumber\\&& \Bigl(\frac{(J-1)(J-4)}{2^{J+1}J} f_{J}(s+1/2)-\frac{2(J-4)%
}{2^{J+1}J} g_{J}(s-1/2) \Bigr) \Bigr)\nonumber \label{sds}
\end{eqnarray}
Here the first line reproduces Eq. (\ref{21}) as it should for the case of vanishing electric field.

Similar to the case of pure magnetic field it is necessary first to
calculate $f_{J}(s)$ and $g_{J}(s)$ for the values of $s$, where these series
are convergent. Finally we must continue analytically the obtained functions of
$s$ to $s=-1/2$. In analogy to the theory of zeta-functions \cite{WW,Bateman},
we shall find integral representations for these functions that are convergent
at all values of $s$ (see Appendix C). Then the resulting integrals can be
evaluated numerically. In addition to representation Eq. (\ref{fpole}) we have the similar one for the function $g_J$ ($s\rightarrow 0$):
\begin{eqnarray}
f_{J}(s\pm1/2) &\approx & \frac{f_{J}^{(-1)}(\pm1/2)}{s} +
f_J^{(0)}(\pm1/2),\quad g_{J}(s\pm 1/2) \approx  \frac{g_{J}^{(-1)}(\pm1/2)}{s}
+ g_J^{(0)}(\pm1/2)
\end{eqnarray}

The calculation of the values of $f_{J}^{(-1,0)}(\pm 1/2),
g_{J}^{(-1,0)}(\pm 1/2)$ is described in Appendix C. It has been shown that
$f_{J}^{(-1)}(\pm 1/2)$ may be nonzero for $J = 2(2K+1), K\in Z$ while
$g_{J}^{(-1)}(- 1/2)$ may be nonzero for $J = 2,4 $. We also have
$f_{J}^{(-1)}(1/2) = 0 $ for $J \ge 3$.
 Remarkably, for $J=2$ the contributions of $f^{(-1)}_{2}(1/2)$ and $g^{(-1)}_{J}(-1/2)$ to the
logarithmic term ($\sim \frac{E^2}{B} \, {\rm log} B$)  cancel each other.
Therefore, we have $l_{eff} = l_{eff}^{(0)} + \Delta l_{eff}$ with
\begin{eqnarray}
-\Delta l_{eff} &=& -\frac{E^{2}}{v B^{J/2}} \gamma^{(0)}_J  ,\nonumber\\
&&\gamma_J =  \Bigl( \frac{1}{4\pi\sqrt{\pi }}g_{s}g_{v}2^{J/2}%
\Gamma(-1/2)\Bigl(\frac{(J-1)(J-4)}{2^{J+2}J}{f^{(0)}_{J}(1/2)}
-\frac{2(J-4)%
}{2^{J+2}J}[{g^{(0)}_{J}(-1/2)}]\Bigr)\Bigr),  \label{sds_J_2_}
\end{eqnarray}

Our results for the effective action at $1\le J \le 10$ are accumulated in
Table \ref{tableseff}. (These results are obtained using the data from Table
\ref{tablefg}). From this table it follows that the $\sim E^2$ correction due to electric
field is decreased fast with the increase of the number of layers. At the same
time, the coefficient before the main term due to the magnetic field is increased.
It is worth mentioning that  at $J=4$ the $\sim E^2$ correction to the effective action
vanishes in this approximation.
The linear term in electric field describes the linear Stark effect, which leads to the spontaneous
(broken symmetry) electric polarization. This term can be presented
as a scalar product
${\bf E} \cdot {\bf P}$, where the vector ${\bf P}$ is directed along
the spontaneous polarization, which in the presence of electric field
is oriented along the field.

\subsection{Conventional regularization}

According to our experience due to the consideration of the case when there is
the magnetic field only, the ultraviolet divergent terms may be present in
 the effective action even if its zeta - regularized version is finite.
Therefore, let us consider the conventional regularization for the suspicious
cases $J=1,2,3$. We subtract the contribution of the fermions at $B=0,E=0$.
However, approaching to this limit is performed along the line of constant
$\lambda$. This means that the boundary conditions for the fermionic fields in
the functional integral for the free fermions are anti - periodic in the
reference frame moving with the velocity $V = \lambda v 2^{1/2} B^{(J-1)/2}$,
($B\rightarrow 0$) along the axis orthogonal to $E$. (Remind that for the
system in the presence of external fields $E$ and $B$ the antiperiodic boundary
conditions are adopted in the reference frame moving with the velocity $E/B$.)

\begin{eqnarray}
-L_{eff}&=&-\frac{g_{s}g_{v}}{2T}\mathrm{Tr}\,\mathrm{log}\,\Bigl(\frac{(i\partial
_{\tau })^{2}+\mathcal{H}^{2}}
{(i\partial _{\tau })^{2}+\mathcal{H}_{0}^{2}}\Bigr)^{} \nonumber\\
 &=& \frac{1}{2T}g_{s}g_{v} \int_{0}^{\infty }dtt^{s-1} \Bigl( \mathrm{Tr}e^{-[(i\partial
_{\tau })^{2}+\mathcal{H}^{2}]t}-\mathrm{Tr}e^{-[(i\partial _{\tau
})^{2}+\mathcal{H}_{0}^{2}]t}\Bigr)|_{s=0} = L_{eff}^{(0)}+\Delta L_{eff}^{(1)}+\nonumber\\
&&+\frac{L_{x}L_{y}}{8\pi\sqrt{\pi }}g_{s}g_{v}\, v (2B)^{1+J/2}\,
\int_{0}^{\infty }dtt^{s-3/2}\Bigl(  \sum_{n\geq J}\,{\rm exp}\Bigl(-(n)_{-J}t
 - t\lambda^2
\frac{(J-4)(2n-(J-1))}{2^{J}J}\Bigr)\nonumber\\&&-\sum_{n\geq J}\,{\rm
exp}\Bigl(-(n)_{-J}t\Bigr) -\int_0^{\infty} d \tau \,{\rm exp} \Bigl(- \tau^{J}
t - t\lambda^2 \frac{(J-4)(2\tau-(J-1))}{2^{J}J}\Bigr)+\int_0^{\infty} d \tau
\,{\rm exp} \Bigl(- \tau^{J} t\Bigr)\Bigr)|_{s=0}
\nonumber\\
\label{detusualG}
\end{eqnarray}
This is the generalization of Eq. (\ref{detusual}) to the case of nonzero $E$.
Here ${\cal H}_0$ is the Hamiltonian of the system at $B,E\rightarrow 0$ but
with the same $\lambda = \frac{E}{v2^{1/2}B^{(J+1)/2}}$ as in $\cal H$.

At $J=1$ Eq. (\ref{detusualG}) is convergent and is equal to the zeta
regularized expression. This means that for the case $J=1$ the zeta -
regularized effective action appears when the contribution of the fermions with
vanishing $B$ and $E$ is subtracted. The limit $E,B\rightarrow 0$ is obtained
along the line of constant $\lambda$.

At $J=2$ there is the logarithmic ultraviolet divergence in Eq.
(\ref{detusualG}). At $J=3$ Eq. (\ref{detusualG}) contains the divergency but
only in  $L_{eff}^{(0)}$.

\subsection{Magnetoelectic effect}

Let us remind (see Appendix A) that
\begin{equation}
-L_{eff}[E,B]=    {F[E,B]} =  -L_x L_y l_{eff}
\end{equation}%
where ${F[E,B]} $ is the free energy of the system calculated in the reference
frame, where electric field vanishes. It is equal to the effective lagrangian
with the minus sign. The quantity
\begin{equation}
\tilde{F} = F + L_x L_y h (B^2 - E^2)/2
\end{equation}
 (where $h$ is the thickness of the graphene sheet) can be considered as the
thermodynamical potential for fixed $E$ and $B$. Actually, $\tilde{F}$ is the
effective lagrangian (with the minus sign) for the system of graphene and the
constant electromagnetic field. As usual, we may introduce vectors $D$ and $H$:
\begin{equation}
D = P + E, \quad H = B - M,
\end{equation}
where $P$ and $M$ are electric and magnetic polarizations (of a unit volume):
\begin{equation}
\frac{dF}{L_x L_y h} = -PdE - M dB,
\end{equation}
Then we have:
\begin{equation}
\frac{1}{L_x L_y h} d \tilde{F} = -D dE + H d B
\end{equation}

Thermodynamical potential $U$ with respect to variables $B$ and $D$ is related
to $\tilde{F}$ as follows:
\begin{equation}
U=\tilde{F}+DE
\end{equation}
Its differential is
\begin{equation}
\frac{dU}{L_x L_y h} = EdD + H dB
\end{equation}
One can see that this is related to the lagrangian $\tilde{L}_{eff} =
-\tilde{F}$ in the same way as the classical Hamiltonian (that is another
definition of energy for the electromagnetic field):
\begin{equation}
U=F-E \frac{\partial F}{\partial E} = E \frac{\partial
\tilde{L}_{eff}}{\partial E} -\tilde{L}_{eff}, \quad \tilde{L}_{eff} = L_{eff}
-L_x L_y h (B^2 - E^2)/2
\end{equation}

For the case of weak electric
field the thermodynamical potential $F$ per unit area depends on $E$ and $B$ as%
\begin{equation}
\frac{1}{L_xL_y}F[E,B]=\frac{1}{L_xL_y} F[E=0,B] - \omega_J B^{1/2}E -
\gamma^{}_J \frac{E^{2}}{v B^{J/2}}
\end{equation}%
where the coefficients are given  in Table \ref{tableseff}.

The magnetization $M=-\frac{1}{L_xL_y h}\frac{\partial F}{\partial B}$ and electric polarization $%
P=-\frac{1}{L_xL_y h}\frac{\partial F}{\partial E}$ of undoped multilayer graphene at
zero temperature can be found from this expression. Due to the effect of
electric field on Landau levels, the magnetoelectric effect arises, that is,
the dependence of magnetization on the electric field and electric polarization
on the
magnetic field. It is characterized by the quantity%
\begin{equation}
h \frac{\partial M}{\partial E}= h \frac{\partial P}{\partial
B}=-\frac{1}{L_xL_y }\frac{\partial ^{2}F}{\partial E\partial
B}=\Bigl(\frac{1}{2}\omega_J B^{-1/2} -\gamma_J J\frac{E}{vB^{\left( J+2\right)
/2}}\Bigr)\label{MAGNETOELECTRIC}
\end{equation}%

It follows from Table \ref{tableseff} that the second term in the
magnetoelectric effect becomes weaker when the number of layers is increased
while the first one is increased with the increase of $J$.

\section{The system in the presence of electric field and small magnetic
field}

\subsection{Semiclassics in one-particle Schr\"odinger equation}
\label{Sectschw}

To establish the relation with the previous work \cite{volovik2012} let us
consider the  gauge $A_{t}=-Ex,A_{y}=Bx$. Then we have stationary
Schr\"odinger equation $H\Psi =\epsilon \Psi $ with
\begin{equation}
H=\left(
\begin{array}{cc}
Ex & v\Bigl(\hat{p}_{x}-i(\hat{p}_{y}+Bx)\Bigr)^{J} \\
v\Bigl(\hat{p}_{x}+i(\hat{p}_{y}+Bx)\Bigr)^{J} & Ex%
\end{array}%
\right)\label{HSectschw}
\end{equation}%
We proceed with the rescaling $z=\Bigl(\frac{E}{v}\Bigr)^{1/(J+1)}x$, $%
\omega =\Bigl(\frac{1}{vE^{J}}\Bigr)^{1/(J+1)}\epsilon $, $\gamma =B\Bigl(%
\frac{v}{E}\Bigr)^{2/(J+1)}$, and $\Pi =\Bigl(\frac{v}{E}\Bigr)^{\frac{1}{J+1%
}}p_{y}$. Then instead of Eq. (\ref{ut}) we have:
\begin{eqnarray}
&&(z-\omega )\psi _{1}+(-i\partial _{z}-i(\Pi +\gamma z))^{J}\psi _{2}=0
\nonumber \\
&&(z-\omega )\psi _{2}+(-i\partial _{z}+i(\Pi +\gamma z))^{J}\psi _{1}=0
\end{eqnarray}%
The first-order semiclassical approximation for $\psi _{1,2}$ gives
\begin{equation}
\psi =e^{\pm i\int (-(\Pi +\gamma z)^{2}+(z-\omega )^{2/J})^{1/2}dz}
\end{equation}%
We denote here $u=(z-\omega )^{2/J}/\Theta ^{2}$ and $\Theta =\Pi +\gamma
\omega $. Then
\begin{equation}
\psi =e^{\mp \Theta ^{J+1}\frac{J}{2}\int ((1+\gamma \Theta
^{J-1}u^{J/2})^{2}-u)^{1/2}u^{J/2-1}du}
\end{equation}%
Integration over the classically forbidden region $(1+\gamma \Theta
^{J-1}u^{J/2})^{2}>u$ gives\ usthe pair production probability. In the limit
$\gamma =0$ ($B=0$) we have
\begin{eqnarray}
|\eta _{0}|^{2} &=&e^{-\alpha _{0}\Theta ^{J+1}},  \nonumber \\
&&\alpha _{0}=2JB\Bigl(\frac{3}{2},\frac{J}{2}\Bigr)
\end{eqnarray}%
The first-order perturbation in $B$ results in \cite{volovik2012}:
\begin{eqnarray}
|\eta |^{2} &=&e^{-\alpha _{0}\Theta ^{J+1}-\gamma ^{2}\alpha _{1}\Theta
^{3J-1}},  \nonumber \\
&&\alpha _{1}=3J^{2}B\Bigl(\frac{1}{2},\frac{3J}{2}\Bigr)
\end{eqnarray}

\subsection{Field-theoretical consideration}

The fact that in our approximation the particles do not interact with each other allows to reduce
the field-theoretical problem to the quantum-mechanical one. Namely, we
arrive at the following pattern. Modes for different values of momenta
propagate independently. At $t\leq t_{0}$ all states with negative values of
energy are occupied while all states with positive values of energy are
vacant. Their evolution in time is governed by the one-particle Schr\"{o}%
dinger equation. At $t=t_{0}+T$ the wave function already has the nonzero
component corresponding to positive energy. Its squared absolute value is
the probability that the electron - hole pair is created. In this section we
imply that the gauge is chosen such that $A_{x}=Et,A_{y}=Bx$. However, the
probability that the electron-hole pair is created at the definite value of
the momentum $p_{y}$ does not depend on the gauge chosen. Therefore we can
use here the results of the previous subsection.

Let us calculate the probability that the vacuum remains vacuum $P_{v}$
(vacuum persistence probability). According to the above presented
calculation this probability is
\begin{equation}
P_{v}=\prod\limits_{p_{x},p_{y}}\Bigl(1-|\eta _{+}|^{2}\Bigr)%
^{g_{s}g_{v}}=e^{-2\mathrm{Im}S}
\end{equation}%
Here $S$ is the effective action, the factors $g_{s}=2$ and $g_{v}=2$ are
spin and valley degeneracies. The product is over the momenta that satisfy
\begin{equation}
ET/2>p_{x}>-ET/2
\end{equation}%
The total probability of the pair creation per unit area per unit time is
\begin{eqnarray}
\omega  &=&\frac{2\,\mathrm{Im}S}{TL_{x}L_{y}}=-g_{s}g_{v}\frac{E}{2\pi L_{x}%
}\sum_{p_{y}=\frac{2\pi }{L_{y}}K}\mathrm{log}(1-|\eta _{+}|^{2})  \nonumber
\\
&\approx &-g_{s}g_{v}\frac{E}{2\pi }\int \frac{dp_{y}}{2\pi }\mathrm{log}%
(1-|\eta _{+}|^{2})  \nonumber \\
&=&g_{s}g_{v}\frac{E^{\frac{J+2}{J+1}}}{2\pi }\Bigl(\frac{1}{v}\Bigr)^{\frac{%
1}{J+1}}\sum_{n}\frac{1}{n}\int \frac{d\Theta }{2\pi }e^{-n\alpha _{0}\Theta
^{J+1}-\gamma ^{2}\alpha _{1}n\Theta ^{3J-1}}
\end{eqnarray}%
The final result reads

\begin{eqnarray}
\omega  &=&g_{s}g_{v}\Bigl(\frac{1}{v}\Bigr)^{\frac{1}{J+1}}\frac{E^{\frac{%
J+2}{J+1}}}{2(J+1)\pi ^{2}(\alpha _{0})^{1/{(J+1)}}}\zeta \Bigl(\frac{J+2}{%
J+1}\Bigr)\Gamma \Bigl(\frac{1}{J+1}\Bigr)  \nonumber \\
&&\Bigl(1-B^{2}\Bigl(\frac{v}{E}\Bigr)^{4/(J+1)}\frac{\alpha _{1}}{\alpha
_{0}^{(3J-1)/(J+1)}}\frac{\zeta \Bigl(\frac{3J}{J+1}\Bigr)\Gamma \Bigl(\frac{%
3J}{J+1}\Bigr)}{\zeta \Bigl(\frac{J+2}{J+1}\Bigr)\Gamma \Bigl(\frac{1}{J+1}%
\Bigr)}\Bigr),  \nonumber \\
&&\alpha _{0}=2JB\Bigl(\frac{3}{2},\frac{J}{2}\Bigr),\quad \alpha
_{1}=3J^{2}B\Bigl(\frac{1}{2},3J/2\Bigr)\label{OmegaEB}
\end{eqnarray}%
According to Ref. \cite{cohen2008} a different quantity is considered as the
pair production rate:
\begin{eqnarray}
\Gamma  &=&\langle |\eta _{+}|^{2}\rangle /(L_{x}L_{y}T)  \nonumber \\
&=&g_{s}g_{v}\Bigl(\frac{1}{v}\Bigr)^{\frac{1}{J+1}}\frac{E^{\frac{J+2}{J+1}}%
}{2(J+1)\pi ^{2}(\alpha _{0})^{1/{(J+1)}}}\Gamma \Bigl(\frac{1}{J+1}\Bigr)%
\Bigl(1-B^{2}\Bigl(\frac{v}{E}\Bigr)^{4/(J+1)}\frac{\alpha _{1}}{\alpha
_{0}^{(3J-1)/(J+1)}}\frac{\Gamma \Bigl(\frac{3J}{J+1}\Bigr)}{\Gamma \Bigl(%
\frac{1}{J+1}\Bigr)}\Bigr) \label{GammaEB}
\end{eqnarray}

\section{Conclusions}
\label{sectconcl}

In this paper we calculated for the first time the effective Euler-Heisenberg action for the multilayer graphene at $ABC$ stacking in the presence of external electric and magnetic fields at $E<< B$. In the opposite limit $B<<E$ we calculated only the imaginary part of the effective action.  The considered effective field model is a kind of the quantum field theory with the anisotropic scaling  ${\bf r}\rightarrow b {\bf r}$, $t\rightarrow b^J t$ that is now
becoming relevant.  The particular anisotropic scaling with $J=3$ has been applied by Ho\v{r}ava for
 construction of the quantum theory of gravity, which does not suffer from the ultraviolet divergences.
Graphene and graphene like materials may serve as the condensed matter realization
of the anisotropic scaling with arbitrary $J$. These materials have  nodes
in the fermionic spectrum, which are characterized by the integer momentum-space topological invariant
$N$. Close to such a node fermions behave as 2+1 massless Dirac particles with energy
spectrum ${\cal E}(p) =\pm v p^N$. These fermions induce the terms in the action for electrodynamic
 fields, which obey  the anisotropic scaling with $J=N$.

Here we considered the fixed space dimension $D=2$, but with point node of arbitrary topological
charge $N$. This is somehow orthogonal to the relativistic systems studied in literature, which
in case of massless fermions corresponds to the fixed $J=1$, but with arbitrary space dimension $D$. In our case some features (in the presence of the external fields $E<<B$) look similar, but some are new:

\begin{enumerate}
\item{} For $D=2$ and $J=2$ the logarithmic term appears that is naturally
expected from
 the one-loop consideration. In relativistic theories the same logarithmic term naturally
 appears in one-loop consideration in conventional electrodynamics with massless fermions, which
   corresponds to $J=1$ and $D=3$.


\item{} For $J \ge 3$ there are terms that are divergent stronger than
logarithmically. These terms depend on  magnetic field but do not depend on
electric field.

\item{} At $J \ge 2$ the magnetoelectric effect is dominated by the lowest
Landau level. Its degeneracy in the absence of electric field is $J$. In the
presence of electric field the degeneracy is eliminated. The corresponding term
in the effective action is proportional to $E$, which produces the analog of the Stark effect. The lowest subdominant terms
are proportional to $E^2$ and are decreased fast with the increase of $J$. There is a specific value
$J=4$, at which the quadratic $E^2$  term is  absent.

\item{} The term  which contains the logarithm $\sim B^{\frac{J+2}{2}} \ln B$ appears only for special values of $J$, such as $J=2,6,10$. This situation is similar to what occurs in relativistic theories, where the existence of the term
$\sim B^{\frac{D+1}{2}} \ln B$ depends on the space dimension $D$ \cite{Visser}. It would be interesting to consider the general case of arbitrary $D$ and $J$.

\end{enumerate}

It is worth mentioning that at $J < 3$ the zeta regularization of the effective action at $E<<B$ gives the correct result. At the same time,
for $J \ge 3$ the zeta regularization gives only the subdominant terms. The dominant ultraviolet divergent terms are calculated as well.
This case is similar to the case of Dirac fermions in the presence of external fields $E<<B$ and with the dimension of  space $D>3$.

In the case $E>>B$ we calculate the imaginary part of the effective action with the small correction $\sim B^2$.

In future it will be instructive to consider electrodynamics arising in general
case of arbitrary space dimension $D$  in the vicinity of the manifold of
zeroes in the fermionic energy spectrum of different dimensions. It appears
that the point nodes in $D=3$  described by topological charge $N>1$ gives rise
to the more complicated structure of the induced electromagnetic action: the
QED has  different scaling laws for different directions in space. For example,
fermions emerging near the Weyl point with topological charge $N=2$ have the
$J=1$ scaling law for spectrum along an anisotropy axis  and the $J=2$ scaling
for the transverse directions (see Sec. 12.4 in Ref. \cite{Volovik2003}).

\section*{Acknowledgements}
This work was partly supported by RFBR grant 11-02-01227, by Grant for
Leading Scientific Schools 6260.2010.2, by the Federal Special-Purpose
Programme 'Cadres' of the Russian Ministry of Science and Education, by
Federal Special-Purpose Programme 07.514.12.4028. MIK acknowledges a
financial support by FOM (the Netherlands). GEV acknowledges a
financial support of the Academy of Finland and its COE program, and the EU � FP7 program ($\#$228464 Microkelvin).

\section*{Appendix A: Effective action and one-particle spectrum}

In expression Eq. (\ref{Z_10}) for the effective action   $\cal H$ is the one -
particle hamiltonian in the presence of external electric and magnetic fields
$E$ and $B$. This hamiltonian may depend on time explicitly (as, for example,
in Eq. (\ref{H1})). { $T \rightarrow \infty$ is time.} It is implied that $E$
is small enough, so the electron - hole pairs are not created.
Our main supposition here is that there exists the transformation $y
\rightarrow \tilde{y} =  y + V t$, with some velocity $V$ such that in new
variables $(x,\tilde{y},t)$ we have  $[i\partial_t - {\cal H}[E,B,t]]\psi(x,y,t)
=[i\partial_t - \tilde{\cal H}[E,B]]\psi(x,\tilde{y},t)$, where $\tilde{H}$ does
not depend on time. For Eq. (\ref{H1}) this is achieved for $V = E/B$.


The system is considered with anti - periodic in time boundary conditions (in
this new coordinates): $\psi(t+T, {x}, \tilde{y}) = - \psi(t, { x},\tilde{y})$.
Suppose we find the solution $\zeta$ of the equation $( i \partial_{t} - {\cal
H} )\zeta = 0$ such that $\zeta_{\cal E}(t + T) = e^{-i {\cal E} T} \zeta_{\cal
E}$. {In this case $\cal E$ is the eigenvalue of $\tilde{\cal H}$ and the
analogue of the energy level. Actually, it is the energy level in the case,
when $\cal H$ does not depend on time from the very beginning, and $V=0$. }
Then $\Psi_{k, {\cal E}} = e^{i\frac{\pi}{T}(2k+1)t + i {\cal E} t}\zeta_{\cal
E}$ is the eigenfunction of the operator $( i
\partial_{t} - {\cal H} ) $:
\begin{equation}
( i \partial_{t} - {\cal H} )\Psi_{k,{\cal E}} = -(\frac{\pi}{T}(2k+1) + {\cal
E})\Psi_{k,{\cal E}}
\end{equation}

The product over $k$ can be calculated as follows
\begin{eqnarray}
\Pi_{ k = - N, ... , N }[\frac{\pi}{T}(2k+1) + {\cal E}] & = &
\Bigl(\Pi_{k=-N,...,N} \frac{\pi}{T}(2k+1)\Bigr) \Bigl(\Pi_{k} [1 + \frac{{\cal
E} T}{\pi(2k+1)}]\Bigr) \nonumber\\ & \approx &
\Bigl(\frac{\pi}{2Na}\Bigr)^{2N} \Bigl(\Pi^N_{k=-N+1} (2k+1)\Bigr) \, {\rm cos}
({\cal E} T/2),
\end{eqnarray}
where $T = 2 N a$, and $a$ is the lattice spacing. This results in
\begin{equation}
{\rm Det} ( i
\partial_{t} - {\cal H}) = e^{-\Omega_0 } \Pi_n {\rm cos} ({\cal E}_n T/2),
\end{equation}
where { $\Omega_0$ depends on the details of the regularization but does not
depend neither on $T$ nor on the spectrum} in continuum limit. The values
${\cal E}_n$ depend on the parameters of the hamiltonian, index $n$ enumerates
these values.

Partition function receives the form (see also \cite{semiclass,rajaraman}):
\begin{eqnarray}
Z(T) & = &   e^{-\Omega_0 }\,\sum_{\{K_n\} = 0,1} \, {\rm exp}\Bigl( i
\frac{T}{2} \sum_n {\cal E}_n - i T \sum_n K_n {\cal E}_n \Bigr)=e^{-\Omega_0
}\,\sum_{\{K_n\} = 0,1} \, {\rm exp}\Bigl(- i T \sum_n K_n {\cal E}_n \Bigr)
 \label{G_p_h_E_}
\end{eqnarray}

Following \cite{semiclass} we interpret Eq. (\ref{G_p_h_E_}) as follows. $K_n$
represents the number of occupied states with the values of "energy" ${\cal
E}_n$, $K_n$ may be $0, 1$. The term $\sum_n {\cal E}_n$ vanishes because the
values ${\cal E}_n$ come in pairs with opposite signs.

After the Wick rotation we arrive at
\begin{eqnarray}
Z(-i \tau) & = &   e^{-\Omega_0 }\,\sum_{\{K_n\} = 0,1} \, {\rm exp}\Bigl(-
\tau \sum_n K_n {\cal E}_n \Bigr)
 \label{G_p_h_E_E}
\end{eqnarray}

Here $\tau = \frac{1}{\cal T}$, { where $\cal T$ is temperature}. In the formal
limit ${\cal T} \rightarrow 0$ only $K_n = [1-{\rm sign} {\cal E}_n]/2$
survives. Thus we get
\begin{eqnarray}
Z(-i/{\cal T}) & = &   e^{-\Omega_0}\, {\rm exp}\Bigl(-
\frac{1}{\cal T} \sum_{n, {\cal E}_n \le 0} {\cal E}_n[E,B] \Bigr), \quad {\cal
T} \rightarrow 0
 \label{G_p_h_E_E_0}
\end{eqnarray}

{ Comparing Eq (\ref{G_p_h_E_E_0}) and Eq. (\ref{Z_10}) we obtain

\begin{eqnarray}
Z({T}) & = &   e^{-\Omega_0}\, {\rm exp}\Bigl(-i  T \sum_{n, {\cal E}_n \le
0} {\cal E}_n[E,B] \Bigr)
 \label{G_p_h_E_E_}
\end{eqnarray}
and
\begin{equation}
L_{eff}[E,B] = - \sum_{n, {\cal E}_n \le 0} {\cal E}_n[E,B] = + (1/2)\sum_{n, }
|{\cal E}_n[E,B]| = - F[E,B]
\end{equation}
Here $F[E,B]$ is the free energy  of the system in the presence of constant
external fields $E$ and $B$ in the reference frame moving with the velocity
$E/B$ in the direction orthogonal to the direction of $E$. (Actually, in this
reference frame the electric field is absent.)   $S = T L_{eff}[E,B]$ is the
effective action.

\section*{Appendix B: Energy levels in the presence of magnetic field and small
electric field}

\subsection{Schrodinger equation}

Here we consider the case when weak in-plane electric field (along the
$x$-axis) is added. We use here the same gauge as in Sect. \ref{Sectschw}. Namely, the external vector potential has the form: $A_{y}=Bx$
while the scalar potential is $A_t = Ex$. The one-particle Hamiltonian  has the
form of Eq. (\ref{HSectschw})
\begin{equation}
H=\left(
\begin{array}{cc}
E x & v\Bigl(\hat{p}_{x}-i(\hat{p}_{y}+Bx)\Bigr)^{J} \\
v\Bigl(\hat{p}_{x}+i(\hat{p}_{y}+Bx)\Bigr)^{J} & Ex \label{H1_}%
\end{array}%
\right)
\end{equation}%
We try the wave function as $\Psi (t,x))=\sum_{p_{y}}e^{ip_{y}y}\psi
_{p}(x,t)$.

\subsection{Zero order approximation}

Let us change the variables:
\begin{equation}
\tilde{\epsilon} = \frac{\mathcal{E}+\frac{E}{B}p_y}{vB^{J/2}},\quad \epsilon =\frac{\mathcal{E}}{vB^{J/2}},\quad\lambda =\frac{E}{\sqrt{2}vB^{(J+1)/2}},\quad u=\sqrt{B}(x+\frac{p_{y}}{B}%
), \quad u_c = -\frac{p_y}{B^{1/2}}
\end{equation}%
and denote
\begin{equation}
\hat{a}^{\dag }=\frac{1}{\sqrt{2}}[-\partial _{u}+u],\quad \hat{a}=\frac{1}{%
\sqrt{2}}[\partial _{u}+u]
\end{equation}%
Instead of Eq. (\ref{ut}) the Schr\"{o}dinger equation reads

\[
\tilde{\epsilon} \psi = (\epsilon-\lambda \sqrt{2} u_c) \psi =\left(
\begin{array}{cc}
\lambda (\hat{a}^{\dag }+\hat{a})  &(-i)^J 2^{J/2}[\hat{a}]^{J} \\
(i)^J2^{J/2}[\hat{a}^{\dag }]^{J} & \lambda (\hat{a}^{\dag }+\hat{a})  %
\end{array}%
\right) \psi
\]%

When $E$ is small, in the zeroth order approximation over $\lambda$ we have:
\begin{eqnarray}
({\cal E}+\frac{E}{B} p_y){\psi }_{1} &=&v(2B)^{J/2}[\hat{a}^{\dag }]^{J}\psi
_{2} \nonumber \\
({\cal E}+\frac{E}{B} p_y){\psi }_{2} &=&v(2B)^{J/2}[\hat{a}]^{J}\psi _{1}
\end{eqnarray}%
(We cannot neglect here $ \frac{E}{B} p_y$ because $p_y$ may be large.) We have
the same Landau levels as without electric field shifted by $ \frac{E}{B} p_y$.

\subsection{The first order correction}

In the next approximation we have
\[
\tilde{\epsilon} \psi =\left(
\begin{array}{cc}
\lambda (\hat{a}^{\dag }+\hat{a}) & (-i)^J2^{J/2}[\hat{a}^{\dag }]^{J} \\
i^J 2^{J/2}[\hat{a}]^{J} & \lambda (\hat{a}^{\dag }+\hat{a})%
\end{array}%
\right) \psi
\]%

One can easily see that the first-order term in expansion of $\epsilon $ in
powers of $\lambda $ vanishes for $n\ge J$. For $n\leq J-1$  there is the first
order correction to the energy for $J\ge 2$. This correction is given by the
eigenvalues $\chi^{(J)}_k \lambda$ of the $J\times J$ matrix
 $\lambda \Omega$
with
\begin{equation}
\Omega = \left(\begin{array}{ccccc}0&1&0&...&...\\
1 & 0 & \sqrt{2} &..&...\\
0&...&...& ...&...\\
0&...&\sqrt{J-2}&0&\sqrt{J-1}\\
0&...&0&\sqrt{J-1}&0
 \end{array}\right)
\end{equation}
We have the same eigenvalues as in Eq. (\ref{chi}).

\subsection{The second order approximation}

One can easily see that the first-order term in expansion of $\epsilon $ in
powers of $\lambda $ vanishes for $n\ge J$ and the second-order term

\begin{equation}
\tilde{\epsilon} _{n}=\tilde{\epsilon} _{n}^{(0)}+\lambda ^{2}\sum_{k\neq n}\frac{1}{%
\tilde{\epsilon} _{n}^{(0)}-\tilde{\epsilon} _{k}^{(0)}}|\psi _{n}^{\dag }(\hat{a}^{\dag }+%
\hat{a})\psi _{k}|^{2} \label{ute_2}
\end{equation}
should be considered. Here we use that
the functions $\psi_n$ are given by Eq. (\ref{psiH}), Eq. (\ref{psiH0}).

For $n\geq J$ Eq. (\ref{ute_2}) can be rewritten as
\begin{eqnarray}
\Delta \tilde{\epsilon} _{n} &=&\lambda ^{2}\eta _{n},\nonumber\\ && \eta_n =
\frac{(J-4)(2n-(J-1))}{2J (n)_{-J}^{1/2} 2^{J/2}} \label{ute_2_}
\end{eqnarray}%

For $n < J$ the second order correction vanishes.

\section*{Appendix C: Calculation of the functions $f_J(s)$ and $g_J(s)$.}

\subsection{Integral representations}

In this section we describe regular procedure for the calculation of $f_J(s)$
and $g_J(s)$.

For $s > 1/J$ ($s> 2/J$) we have
\begin{eqnarray}
 f_J(s) &=& \sum_{n\ge J} \frac{1}{[(n)_{-J}]^s}, \quad g_J(s) = \sum_{n\ge J}
 \frac{n}{[(n)_{-J}]^{s+1}}\label{series}
\end{eqnarray}

We apply to these sums the Plana summation formula \cite{Bateman}, Vol. 1,
1.9(11):

\begin{eqnarray}
\sum_{n=0}^{\infty} f(n) & = & \frac{1}{2} f(0)  + \int^{\infty}_{0} f(\tau)
\,d\tau + i \int^{\infty}_{0}\Bigl(f(i t) - f( - i t)\Bigr)\frac{dt}{e^{2 \pi
t} - 1}
\end{eqnarray}

This formula  works if:

\begin{enumerate}
\item{} $f(x)$ is regular for ${\rm Re}\, x \ge 0$,

\item{} $e^{-2\pi |t|}f(\tau + it) \rightarrow 0$ at $t \rightarrow \infty$,
$0\le \tau < \infty$,

\item{} $\int e^{-2\pi |t|}|f(\tau + it)| dt \rightarrow 0$ at $\tau
\rightarrow \infty$
\end{enumerate}

Therefore, we obtain:
\begin{eqnarray}
f_J(s) & = & \frac{1}{2 [J!]^s}  + \int^{\infty}_{0} (\tau + J)_{-J}^{-s}
\,d\tau
+ i \int^{\infty}_{0}\Bigl((i t + J)_{-J}^{-s} - ( - i t + J)_{-J}^{-s}\Bigr)\frac{dt}{e^{2 \pi t} - 1}, \quad s > 1/J\nonumber\\
g_J(s) & = & \frac{J}{2 [J!]^s}  + \int^{\infty}_{0} (\tau + J)_{-J}^{-s-1}
(\tau+J) \,d\tau \nonumber\\ &&+ i \int^{\infty}_{0}\Bigl((i t +
J)_{-J}^{-s-1}( i t + 1) - (-i t + J)_{-J}^{-s-1}(- i t +
1)\Bigr)\frac{dt}{e^{2 \pi t} - 1},\quad s
> 2/J
\end{eqnarray}
Here
\begin{eqnarray}
(\tau + J)_{-J} & = & \frac{\Gamma(\tau + J + 1)}{\Gamma(\tau + 1)} = (\tau +
J) (\tau+J-1) ... (\tau  + 1)\label{s+}
\end{eqnarray}

Now the divergences of the sums over $n$  at $s \le 1/J$ and $s \le 2/J$
correspondingly are concentrated within the integrals over $\tau$. At $s > 1/J$
($s> 2/J$) we may represent these integrals as follows:
\begin{eqnarray}
&& I_J(s)  =  \int^{\infty}_{0} (\tau + J)_{-J}^{-s} \,d\tau, \quad
\tilde{I}_J(s)
=  \int^{\infty}_{0} (\tau + J)_{-J}^{-s-1} (\tau+J) \,d\tau \nonumber\\
&& 0 = \oint_C z_{-J}^{-s} dz = I_J(s)(1+e^{-i\pi Js}) + \int_{-1}^{J}
z_{-J}^{-s} dz
\nonumber\\
&& 0 = \oint_C z_{-J}^{-s-1} z dz = \tilde{I}_J(s)(1-e^{-i\pi J(s+1)}) + (J-1)
e^{-i \pi J(s+1)} I_J(s+1) + \int_{-1}^{J} z_{-J}^{-s-1} z dz
\end{eqnarray}

Here contour $C$ consists of the integral from $-\infty+i0$ to $-1+i0$, the
part of the circle that belongs to the upper half - plane, starts at $-1+i0$,
and ends at $J + i0$, then the part of the real axis from $J+i0$ to
$\infty+i0$. The contour is closed via the half - circle at infinity in the
upper half - plane. We obtain:
\begin{eqnarray}
&& I_J(s)  = -\frac{1}{1+e^{-i\pi Js}}\oint_{-1}^{J} z_{-J}^{-s} dz \nonumber\\
&& \tilde{I}_J(s) = -\frac{1}{1-e^{-i\pi J(s+1)}} \oint_{-1}^{J} z_{-J}^{-s-1}z
dz - \frac{(J-1)e^{-i\pi J(s+1)}}{1-e^{-i\pi J(s+1)}} I_{J}(s+1)\label{II}
\end{eqnarray}
Here integrals $\oint$ are taken along the half - circle $z = \frac{J-1}{2} +
\frac{J+1}{2} e^{i\omega},\, \omega \in [0, \pi] $. The given integrals are
convergent for all values of $s$, probably, except for $s = (2K+1)/J$ and $s =
2Q/J -1$, $K,Q\in Z$.

For the ordinary Riemann zeta function there is the Hermit representation
(\cite{Bateman}, Vol.1, 1.10, (7)). Using the expressions listed above we
derive the analogue of this representation:
\begin{eqnarray}
f_J(s) & = & \frac{1}{2 [J!]^s}  -\frac{1}{1+e^{-i\pi Js}}\oint_{-1}^{J}
z_{-J}^{-s} dz  + 2 \int^{\infty}_{0}\frac{{\rm sin} \, s [{\rm arctg}\,
\frac{t}{J} + ... +{\rm arctg}\,t] }{(t^2+J^2)^{s/2}...(t^2+1)^{s/2}}
\frac{dt}{e^{2 \pi t} - 1}, \nonumber\\
g_J(s) & = & \frac{J}{2 [J!]^s}  - \frac{1}{1-e^{-i\pi J(s+1)}} \oint_{-1}^{J}
z_{-J}^{-s-1}z dz  + \frac{J-1}{1-e^{-i\pi J(s+1)}}
\frac{e^{-i\pi J(s+1)}}{1+e^{-i\pi J(s+1)}}\oint_{-1}^{J} z_{-J}^{-s-1} dz\nonumber\\
&&+ 2 \int^{\infty}_{0}\frac{{\rm sin}( \, (s+1) [{\rm arctg}\, \frac{t}{J} +
... +{\rm arctg}\,t] - {\rm arctg}\,\frac{t}{J}])
}{(t^2+J^2)^{(s+1)/2}...(t^2+1)^{(s+1)/2}  } \frac{(t^2+J^2)^{1/2}\, dt}{e^{2
\pi t} - 1}\label{Hermit}
\end{eqnarray}
This expression gives the analytical continuation of the series Eq.
(\ref{series}) to all values of $s$. From this representation we conclude that
$f_J(s)$ may have a simple pole at $s=(2K+1)/J, K\in Z$ while $g_J(s)$ may have
poles at $s=(2K+1)/J$ or $s=(2Q+1)/J - 1$.

\subsection{Particular cases}
At $J=1$ expression (\ref{Hermit}) is the Hermit integral representation for
the $\zeta$ - function (see \cite{Bateman}, Vol.1, 1.10, (7)).

We have ($s\rightarrow 0$):
\begin{eqnarray}
f_{J}(s\pm1/2) &\approx & \frac{f_{J}^{(-1)}(\pm1/2)}{s} +
f_J^{(0)}(\pm1/2),\quad g_{J}(s\pm 1/2) \approx  \frac{g_{J}^{(-1)}(\pm1/2)}{s}
+ g_J^{(0)}(\pm1/2)
\end{eqnarray}
We have verified integral representations Eq. (\ref{Hermit}) using MAPLE
package  for several values of $s$ and $J$ such that the series Eq.
(\ref{series}) are convergent. Next, we calculated numerically the values of
$f^{(0,-1)}_J(\pm 1/2)$ and $g^{(0,-1)}_J(-1/2)$ using these integral
representations at $J = 2,..., 10$.

At $J = 2$ we come to:
\begin{eqnarray}
f_2(s) & = & \frac{1}{2^{s+1}}  + I_2(s) + 2 \int^{\infty}_{0}\frac{{\rm sin}
\, s [{\rm arctg}\, \frac{t}{2} + {\rm arctg}\,t]
}{(t^2+2^2)^{s/2}(t^2+1)^{s/2}}
\frac{dt}{e^{2 \pi t} - 1}, \nonumber\\
g_2(s) & = & \frac{1}{2^{s}}  + \tilde{I}_2(s)  + 2 \int^{\infty}_{0}\frac{{\rm
sin}( \, (s+1) [{\rm arctg}\, \frac{t}{2} +{\rm arctg}\,t] - {\rm
arctg}\,\frac{t}{2}]) }{(t^2+2^2)^{(s+1)/2}(t^2+1)^{(s+1)/2} }
\frac{(t^2+2^2)^{1/2}\, dt}{e^{2 \pi t} - 1}\label{Hermit2}
\end{eqnarray}

For $J=2$ we may check expression Eq. (\ref{II}) using  integral representation
for the hypergeometric function (\cite{Bateman}, (2.12), Eq. (5)):
\begin{eqnarray}
&& I_2(s)  = \frac{1}{2^{s+1}(s-1/2)} F(s,1,2s,1/2) = \frac{1}{2^{s+1}(s-1/2)}\Bigl(1+\frac{s}{2s}\frac{1}{2}+\frac{s(s+1)}{2s(2s+1)}\frac{1}{2^2} +...  \Bigr) \nonumber\\
&& \tilde{I}_2(s) = \frac{1}{2^{s+1}s} F(s,1,2s+1,1/2) =
\frac{1}{2^{s+1}s}\Bigl(1+\frac{s}{2s+1}\frac{1}{2}+\frac{s(s+1)}{(2s+1)(2s+2)}\frac{1}{2^2}
+...  \Bigr),
\end{eqnarray}

Therefore, we calculate:
\begin{eqnarray}
 f^{(0)}_2(-\frac{1}{2}) & = & \frac{1}{\sqrt{2}}   + \partial_s[\frac{s+1/2}{2^{s+1}(s-1/2)} F(s,1,2s,1/2)]|_{s=-1/2} \nonumber\\ &&
+ 2 \int^{\infty}_{0}\frac{{\rm sin} (-(1/2)({\rm arctg}\, (t/2) + {\rm arctg}\, t )) }{(4+t^2)^{-1/4}(1+t^2)^{-1/4}}\,  \frac{dt}{e^{2 \pi t} - 1} \nonumber\\
&=& -0.3321609172 \nonumber\\
 f^{(-1)}_2(-\frac{1}{2}) & = &   -
\frac{1}{8\sqrt{2}}(\frac{1}{4} +
\frac{3!!}{1!}\frac{1}{4^2} + \frac{5!!}{2!}\frac{1}{4^3} + ...) =  -0.0625,\nonumber\\
f_2^{(0)}(\frac{1}{2}) & = & \frac{1}{2\sqrt{2}}
+\partial_s[\frac{1}{2^{s+1}} F(s,1,2s,1/2)]|_{s=1/2}  \nonumber\\
&&
+ 2 \int^{\infty}_{0}\frac{{\rm sin} ((1/2)({\rm arctg}\, (t/2) + {\rm arctg}\, t )) }{(4+t^2)^{1/4}(1+t^2)^{1/4}}\,  \frac{dt}{e^{2 \pi t} - 1}\nonumber\\
&=& 0.01902854064, \nonumber\\
f^{(-1)}_2(\frac{1}{2}) & = &   \frac{1}{2^{3/2}}F(1/2,1,1,1/2)=1/2,\nonumber\\
g^{(-1)}_2(-\frac{1}{2}) & = &  \frac{1}{2^{5/2}}F(1/2,1,1,1/2)=1/4,\nonumber\\
g_2^{(0)}(-\frac{1}{2}) & = & {\sqrt{2}}  + \partial_s[\frac{s+1/2}{2^{s+1}s} F(s,1,2s+1,1/2)]|_{s=-1/2} \nonumber\\
&& + 2 \int^{\infty}_{0}\frac{{\rm sin} ((1/2)(-{\rm arctg}\, (t/2) + {\rm
arctg}\, t )) }{(4+t^2)^{-1/4}(1+t^2)^{1/4}}\,  \frac{dt}{e^{2 \pi t} -
1}\nonumber\\ &=&- 0.8676318824, \label{J1}
\end{eqnarray}

For $J > 2$ we proceed in a different way. For example, for the case $J = 3$ we
have
\begin{eqnarray}
f_3(s) & = & \frac{1}{2}\frac{1}{6^{s}}  + I_3(s) + 2
\int^{\infty}_{0}\frac{{\rm sin} \, s [{\rm arctg}\, \frac{t}{3} + {\rm
arctg}\, \frac{t}{2}  +{\rm arctg}\,t]
}{(t^2+3^2)^{s/2}(t^2+2^2)^{s/2}(t^2+1)^{s/2}}
\frac{dt}{e^{2 \pi t} - 1}, \nonumber\\
g_3(s) & = & \frac{1}{2}\frac{3}{6^{s+1}}  + \tilde{I}_3(s)  + 2
\int^{\infty}_{0}\frac{{\rm sin}( \, (s+1) [{\rm arctg}\, \frac{t}{3} + {\rm
arctg}\, \frac{t}{2}  +{\rm arctg}\,t] - {\rm arctg}\,\frac{t}{3}])
}{(t^2+3^2)^{(s+1)/2}(t^2+2^2)^{(s+1)/2}(t^2+1)^{(s+1)/2} }
\frac{(t^2+3^2)^{1/2}\, dt}{e^{2 \pi t} - 1}\label{Hermit3}
\end{eqnarray}
Here $I_3(s)$ and $\tilde{I}_3(s)$ are to be calculated using Eq. (\ref{II}).
We get:
\begin{eqnarray}
 f^{(0)}_3(-\frac{1}{2}) & = & 0.1580218749, \quad  f_3^{(0)}(\frac{1}{2})  =  1.687512381, \quad g_3^{(0)}(-\frac{1}{2})  =   -0.576826000, \label{J3}
\end{eqnarray}

\begin{table}
\begin{center}
\begin{small}
\begin{tabular}{|c|c|c|c|c|c|c|c|c|c|c|c|c|c|c|c|c|}
\hline
$J$ & $f^{(0)}_J(-\frac{1}{2})$ & $f^{(-1)}_J(-\frac{1}{2})$ & $f^{(0)}_J(\frac{1}{2})$ & $f^{(-1)}_J(\frac{1}{2})$ & $g^{(0)}_J(-\frac{1}{2})$ & $g^{(-1)}_J(-\frac{1}{2})$   \\
\hline $1$ & $-0.2078862251$ & $0$ & $-1.460354509$ & $0$ &$-0.2078862251$& $0$\\
\hline
$2$ & $-0.3321609172$ & $-0.0625$ &$0.01902854064 $& $ 0.5$& $ - 0.8676318824$ & $ 0.25$\\
\hline
$3$ & $0.1580218749$ & $0$ & $1.687512381 $ & $0$ & $-0.576826000 $ & $ 0$ \\
\hline $4$ & $ -0.2742668544$ & $0$ & $0.5508597242$ & $0$ & $ 0.3077772742$ & $0.25$ \\
\hline
$5$ & $-1.650991958$ & $0$ & $0.2018069599 $ & $0$ & $1.792805053$ & $ 0$ \\
\hline
$6$ & $-1.299466810$ & $-0.2460937500$ & $0.07309406118$ & $0$ & $0.5888090809$ & $ 0$ \\
\hline
$7$ & $1.789507452$ & $0$ & $0.02547240488$ & $0$ & $0.2149754417 $ & $ 0$ \\
\hline
$8$ & $-10.74171499$ & $0$ & $0.008486905628$ & $0$ & $0.07746594906 $ & $ 0$ \\
\hline
$9$ & $-73.39264066$ & $0$ & $0.002702793021 $ & $0$ & $0.02686060136$ & $ 0$ \\
\hline
$10$ & $-28.2691703$ & $-7.922363278$ & $0.0008241640781$ & $0$ & $0.008909168673$ & $ 0$ \\
\hline
\end{tabular}
\end{small}
\end{center}
\caption{The values of $f_J(\pm 1/2)$ and $g_J(-1/2)$. } \label{tablefg}
\end{table}

At $J = 6,10$ the residues $f^{(-1)}(-1/2)$ are calculated as follows:
\begin{eqnarray}
f^{(-1)}_J(-1/2) & = &   -\frac{1}{i\pi J} \oint_{-1}^{J} z_{-J}^{1/2} dz
\end{eqnarray}
The values $f^{(0)}_J(-1/2)$ are calculated as
\begin{eqnarray}
f^{(0)}_J(-1/2) & = & \frac{1}{2 [J!]^{-1/2}}
-\partial_s\frac{s+1/2}{1+e^{-i\pi Js}}\oint_{-1}^{J} z_{-J}^{-s}
dz|_{s=-1/2}\nonumber\\&& + 2 \int^{\infty}_{0}\frac{{\rm sin} \, (-1/2) [{\rm
arctg}\, \frac{t}{J} + ... +{\rm arctg}\,t]
}{(t^2+J^2)^{-1/4}...(t^2+1)^{-1/4}} \frac{dt}{e^{2 \pi t} - 1}
\end{eqnarray}

 In a
similar way we have calculated the other values collected in Table
\ref{tablefg}. For $J > 4$ the original expression for $g_J(-1/2)$ Eq.
(\ref{series}) is convergent, and we use it to obtain the corresponding values.
Therefore, $g_J^{(-1)}(-1/2) = 0, J
> 4$. In a similar way $f^{}_J(1/2)$ is convergent for $J>2$ and
$f_J^{(-1)}(1/2)=0, J
>2$. At the same time it is not excluded that $f^{(-1)}_J(-1/2) \ne 0$ at $J =
6,10,14,...$. Indeed, we calculate $f^{(-1)}_J(-1/2)$ for $J =6,10$ and find
that these values do not vanish.

\section*{Appendix D: Calculation of the ultraviolet divergent terms in $F(-1/2)$.}

\subsection{The case $J=2$}

For $J=2$ we have:
\begin{eqnarray}
 F_2(s,\delta)& = &
\frac{1}{\Gamma(s)}\int_{\delta}^{\infty }dtt^{s-1}\Bigl( \sum_{n\geq
2}\,e^{-\frac{n!}{(n-2)!}t}+2/2-\int_0^{\infty} d \tau \,e^{- \tau^{2}
t}\Bigr)\label{detusual2}
\end{eqnarray}
with $\delta = \frac{(2vB)^2}{\Lambda^2}$, where $\Lambda$ is the ultraviolet
cutoff. We rewrite this expression using Plana's summation formula as follows:
\begin{eqnarray}
 F_2(s,\delta)& = &
\frac{1}{\Gamma(s)}\int_{\delta}^{\infty }dtt^{s-1}\Bigl( \sum_{n\geq
2}\,e^{-\frac{n!}{(n-2)!}t}+2/2-\int_0^{\infty} d \tau \,e^{- \tau^{2}
t}\Bigr)\nonumber\\
&=&\frac{1}{\Gamma(s)}\int_{\delta}^{\infty
}dtt^{s-1}\Bigl(\frac{1}{2}[e^{-2!t}-1]+ \int_0^{\infty} d\tau
\,e^{-(\tau+1)(\tau+2)t}+3/2-\int_0^{\infty} d \tau \,e^{- \tau^{2}
t}\Bigr)\nonumber\\
&& +  {2} \int^{\infty}_{0}\frac{{\rm sin} \, s [{\rm arctg}\, \frac{t}{2} +
{\rm arctg}\,t] }{(t^2+2^2)^{s/2}(t^2+1)^{s/2}} \frac{dt}{e^{2
\pi t} - 1}\nonumber\\
&=&\frac{1}{\Gamma(s)}\int_{\delta}^{\infty
}dtt^{s-1}\Bigl(3/2+\frac{1}{t^{1/2}}\int_0^{\infty} d\tau
\,[e^{-(\tau+t^{1/2})(\tau+2t^{1/2})}-e^{- \tau^{2}
}]\Bigr)\nonumber\\
&&+\frac{1}{2(2!)^s}+{2} \int^{\infty}_{0}\frac{{\rm sin} \, s [{\rm arctg}\,
\frac{t}{2} + {\rm arctg}\,t] }{(t^2+2^2)^{s/2}(t^2+1)^{s/2}} \frac{dt}{e^{2
\pi t} - 1},\label{detusual2_}
\end{eqnarray}

At $ s=-1/2$ the second integral over $t$ is convergent while in the first one
there is the divergent term that corresponds to the integration over the region
of small $t$:
\begin{eqnarray}
 F_2^{div}(-1/2,\delta)& \approx &
\frac{1}{\Gamma(-1/2)}\int_{\delta}^{...}dtt^{-1/2-1}\Bigl( \frac{1}{8} t^{1/2}
\Gamma(1/2)\Bigr) \approx \frac{1}{16}\, {\rm log}\,  \delta
\end{eqnarray}
This expression gives the logarithmic term in the effective action after the
renormalization of effective charge (that results in the change $\Lambda
\rightarrow \mu$). At the same time at $-1/2<s<0$ expression Eq.
(\ref{detusual2_}) is convergent, but has the pole at $s=-1/2$ with the residue
$-1/16$. In this region we have:
\begin{eqnarray}
 {F}_2(s,0)& = &\frac{1}{\Gamma(s)[1-e^{2\pi i s}]}\oint_{\infty}^{0^+}dtt^{s-1}\Bigl(3/2+\frac{1}{t^{1/2}}\int_0^{\infty} d\tau
\,[e^{-(\tau+t^{1/2})(\tau+2t^{1/2})}-e^{- \tau^{2}
}]\Bigr)\nonumber\\
&&+\frac{1}{2(2!)^s}+{2} \int^{\infty}_{0}\frac{{\rm sin} \, s [{\rm arctg}\,
\frac{t}{2} + {\rm arctg}\,t] }{(t^2+2^2)^{s/2}(t^2+1)^{s/2}} \frac{dt}{e^{2
\pi t} - 1},\label{detusual2__}
\end{eqnarray}
Here integral $\oint_{\infty}^{0^+}$ is over the closed path that starts at
infinity, goes to zero along the real axis, then encloses $t =0$ clockwise and
comes back to infinity along the real axis. This contour can be deformed in
such a way that it is placed at infinity and is wrapped around $t = 0$
clockwise. Written in this form Eq. (\ref{detusual2__}) is convergent for all
values of $s$ except for $s = 0, \pm 1, \pm 2,...$. For $s>1/2$ we come to
expression of Eq. (\ref{Hermit2}) and thus derive $F_2(s,0) = f_2(s), s > -1/2$
.

\subsection{The case $J=3$}

 For $J=3$ we have:
\begin{eqnarray}
 F_3(s,\delta)& = &
\frac{1}{\Gamma(s)}\int_{\delta}^{\infty }dtt^{s-1}\Bigl( \sum_{n\geq
2}\,e^{-\frac{n!}{(n-3)!}t}+3/2-\int_0^{\infty} d \tau \,e^{- \tau^{3}
t}\Bigr)\label{detusua3}
\end{eqnarray}
with $\delta = \frac{v^2(2B)^3}{\Lambda^2}$, where $\Lambda$ is the ultraviolet
cutoff. We rewrite this expression using Plana's summation formula as follows:
\begin{eqnarray}
F_3(s,\delta) &=&\frac{1}{\Gamma(s)}\int_{\delta}^{\infty }dtt^{s-1}\Bigl(
\frac{1}{t^{1/3}}\int_0^{\infty} d\tau
\,e^{-(\tau+t^{1/3})(\tau+2t^{1/3})(\tau+3t^{1/3})}+4/2-\frac{1}{t^{1/3}}\int_0^{\infty}
d \tau \,e^{- \tau^{3}
}\Bigr)\nonumber\\
&& +\frac{1}{2 (3!)^s}+  {2}\int^{\infty}_{0}\frac{{\rm sin} \, s [{\rm
arctg}\, \frac{t}{3}+{\rm arctg}\, \frac{t}{2} + {\rm arctg}\,t]
}{(t^2+3^2)^{s/2}(t^2+2^2)^{s/2}(t^2+1)^{s/2}} \frac{dt}{e^{2 \pi t} -
1},\label{detusua3_}
\end{eqnarray}

The second integral over $t$ is convergent while in the first one there is the
divergent term
\begin{eqnarray}
 F_3^{div}(-1/2,\delta)& = &
\frac{1}{\Gamma(-1/2)}\int_{\delta}^{\infty}dtt^{-1/2-1}\Bigl(
\frac{1}{3}\Gamma(2/3) t^{1/3}\Bigr) \approx -0.7639792361
\,\Bigl(\frac{v^2(2B)^3}{\Lambda^2}\Bigr)^{-1/6}
\end{eqnarray}
This  gives the divergent term in the effective action $\sim v
(2B)^{1+3/2}\Bigl(\frac{v^2(2B)^3}{\Lambda^2}\Bigr)^{-1/6}\sim B^2
\Lambda^{1/3} v^{2/3}$.

\begin{table}
\begin{center}
\begin{small}
\begin{tabular}{|c|c|c|c|c|c|c|c|c|c|c|c|c|c|c|c|c|}
\hline
$J$ &{\bf  } $F_J^{div}(-1/2,\delta)$   \\
\hline
$1$ & $-$ \\
\hline $2$ & $\frac{1}{16}\, {\rm log}\,  \delta$ \\
\hline $3$ & $-0.7639792361
\,\delta^{-1/6}$ \\
\hline
$4$ & $-0.8642091734\, \delta^{-1/4}$\\
\hline
$5$ & $-1.094743796 \, \delta^{-3/10} $  \\
\hline $6$ & $ -1.393109122 \, \delta^{-1/3} + 0.2460937500 \, {\rm log}\,  \delta$  \\
\hline
$7$ & $-1.746814339 \, \delta^{-5/12} -6.155346286 \,\delta^{-1/12}$  \\
\hline
$8$ & $ -2.151696783 \, \delta^{-3/8} -5.842427394 \,\delta^{-1/8}$  \\
\hline $9$ & $-2.605972745 \times
\delta^{-7/18}-6.875813125\times \delta^{-1/6}$ \\
\hline $10$ & $-3.108750471 \times
\delta^{-2/5}-8.590800550 \times \delta^{-1/5} + 7.922363278\, {\rm log}\,  \delta$ \\
\hline
\end{tabular}
\end{small}
\end{center}
\caption{Divergent terms in $F_J(-1/2,\delta)$. Here  $\delta =
\frac{v^2(2B)^J}{\Lambda^2}$, where $\Lambda$ is the ultraviolet cutoff.  }
\label{tabledelta}
\end{table}

\subsection{Arbitrary $J$}

For arbitrary $J$ we have:
\begin{eqnarray}
 F_J(s,\delta)& = &
\frac{1}{\Gamma(s)}\int_{\delta}^{\infty }dtt^{s-1}\Bigl( \sum_{n\geq
2}\,e^{-\frac{n!}{(n-J)!}t}+J/2-\int_0^{\infty} d \tau \,e^{- \tau^{J}
t}\Bigr)\label{detusua3}
\end{eqnarray}
with $\delta = \frac{v^2(2B)^J}{\Lambda^2}$, where $\Lambda$ is the ultraviolet
cutoff. We rewrite this expression using Plana's summation formula as follows:
\begin{eqnarray}
F_J(s,\delta) &=&\frac{1}{\Gamma(s)}\int_{\delta}^{\infty }dtt^{s-1}\Bigl(
\frac{1}{t^{1/J}}\int_0^{\infty} d\tau
\,e^{-(\tau+t^{1/J})...(\tau+Jt^{1/J})}+(J+1)/2-\frac{1}{t^{1/J}}\int_0^{\infty}
d \tau \,e^{- \tau^{J}
}\Bigr)\nonumber\\
&& +\frac{1}{2 (J!)^s}+  {2}\int^{\infty}_{0}\frac{{\rm sin} \, s [{\rm
arctg}\, \frac{t}{J}+... + {\rm arctg}\,t] }{(t^2+J^2)^{s/2}...(t^2+1)^{s/2}}
\frac{dt}{e^{2 \pi t} - 1},\label{detusua3_}
\end{eqnarray}

The second integral over $t$ is convergent while in the first one there may
appear the divergent terms
\begin{eqnarray}
 F_J^{div}(-1/2,\delta)& = &
\frac{1}{\Gamma(-1/2)}\int_{\delta}^{\infty}dtt^{-1/2-1}\Bigl(\sum_{1\le K\le
J/2} A_K t^{K/J}\Bigr) \nonumber\\&\approx& \sum_{1\le K< J/2} \tilde{A}_K\,
\delta^{-(J-2K)/(2J)}+\tilde{A}_{J/2}\, {\rm log}\, \delta
\end{eqnarray}
Here the $\rm log$ term may exist only for even $J$. The divergent  terms for
$1\le J \le 10$ are represented in Table \ref{tabledelta}.

\end{document}